\documentclass[12pt,a4paper]{article}

\usepackage[utf8]{inputenc}
\usepackage{mathpazo}
\usepackage{amssymb}
\usepackage{amsmath}
\usepackage{mathtools}
\usepackage[dvipsnames]{xcolor}
\usepackage[colorlinks]{hyperref}

\hypersetup{
 linkcolor=RoyalPurple
,citecolor=NavyBlue
}

\newcommand{\nc}{\newcommand}
\newcommand{\rnc}{\renewcommand}

\rnc{\baselinestretch}{1.24}	
\rnc{\arraystretch}{1.24}   	

\makeatletter
\rnc{\theequation}{\thesection.\arabic{equation}}
\@addtoreset{equation}{section}
\makeatother                      

\nc{\eq}[1]{(\ref{#1})}
\nc{\newcaption}[1]{\centerline{\parbox{6in}{\caption{#1}}}}

\nc{\fig}[3]{
\begin{figure}
\centerline{\epsfxsize=#1\epsfbox{#2.eps}}
\newcaption{#3. \label{#2}}
\end{figure}
}

\def\l{\lambda}

\def\l({(\!(}
\def\r){)\!)}

\begin{document}

\begin{titlepage}

\begin{flushright}
\end{flushright}
\vspace*{2.0cm}
\centerline{\Large\bf Classical observables from partial wave amplitudes}
\vspace*{1.5cm} 
\centerline{Hojin Lee${}^a$, Sangmin Lee${}^{a,b,c,d}$, Subhajit Mazumdar${}^b$}
\vspace*{1.0cm}
\centerline{\sl ${}^a$Department of Physics, Seoul National University, Seoul 08826 Korea} 
\centerline{\sl ${}^b$Center for Theoretical Physics, Seoul National University, Seoul 08826 Korea}
\centerline{\sl ${}^c$College of Liberal Studies, Seoul National University, Seoul 08826 Korea}
\centerline{\sl ${}^d$School of Physics, Korea Institute for Advanced Study, Seoul 02455 Korea}

\vspace*{2.0cm}
\centerline{\bf ABSTRACT}
\vspace*{0.5cm}
\noindent

We study the formalism of Kosower-Maybee-O'Connell (KMOC) to extract classical impulse from quantum amplitude in the context of the partial wave expansion of a 2-to-2 elastic scattering. 
We take two complementary approaches to establish the connection. 
The first one takes advantage of Clebsch-Gordan relations for the base amplitudes of the partial wave expansion. The second one is a novel adaptation of the traditional saddle point approximation in the semi-classical limit. 
In the former, an interference between the S-matrix and its conjugate leads to a large degree of cancellation such that the saddle point approximation to handle a rapidly oscillating integral is no longer needed. 
As an example with a non-orbital angular momentum, we apply our methods to the charge-monopole scattering problem in the probe limit and reproduce both of the two angles characterizing the classical scattering. 
A spinor basis for the partial wave expansion, a non-relativistic avatar of the spinor-helicity variables, plays a crucial role throughout our computations. 


\vspace*{2.0cm}


\end{titlepage}
\setcounter{footnote}{0}

\tableofcontents


\newpage 
\section{Introduction}

One of the most fruitful applications of the modern technology for scattering amplitudes concerns the dynamics of two massive bodies 
that are bound to each other by gravity and are producing gravitational radiation. 
The amplitude-based methods supplement more traditional methods 
and are pushing the boundaries of post-Newtonian/Minkowskian expansions 
to an extent that was inconceivable a decade ago. 
See \cite{Bjerrum-Bohr:2022blt,Kosower:2022yvp,Buonanno:2022pgc} and references therein for an overview of 
this rapidly unfolding subject. 

The ``classical gravity from quantum amplitudes" program is broadly divided into two parts: the conservative sector and the radiative sector. 
In the former, which will be the focus of this paper, the two body system is described by an effective Hamiltonian accounting for the gravitational interaction. 
The physical observables computed from the Hamiltonian (or sometimes extracted directly from amplitudes) 
include momentum transfer (impulse), spin kick, periastron precession and time delay.

The main goal of this paper is to establish a connection between the partial wave expansion of a 2-to-2 elastic scattering and the method of Kosower-Maybee-O'Connell (KMOC) \cite{Kosower:2018adc} to compute the classical impulse from a quantum amplitude. 
The KMOC method has been combined with other techniques and found numerous applications \cite{Kosower:2022yvp}, but to the best of our knowledge, its connection to the partial wave expansion has not been explored in depth. 

We propose two complementary approaches to establish the connection. 
As we will review in section~\ref{sec:KMOC-review}, in the center of momentum (CM) frame, the KMOC impulse formula 
can be written as an integral: 
\begin{align}
\begin{split}
         \Delta \vec{k}  &= \frac{\hbar^2}{|\vec{k}|^2} \int \hat{d}^2 q \,e^{i \vec{q}\cdot \vec{b} } \,\vec{K} (\vec{k},\vec{q}) \,,
         \\
         \vec{K}(\vec{k},\vec{q})  &= \int \hat{d}^2 \hat{k}' \left[ S^\dagger(\vec{k} + \hbar \vec{q}/2,\vec{k}') (\vec{k}' - \vec{k}) S(\vec{k}', \vec{k} - \hbar \vec{q}/2) \right] \,.
\end{split}
\label{KMOC-full-CM-copy} 
\end{align}
The S-matrix $S$ and its conjugate $S^\dagger$ are to be given as partial wave sums. Their building blocks, ``base amplitudes'', are spinor products to be introduced in section~\ref{sec:partial-wave}. 

In section~\ref{sec:KMOC-PW}, we present the two ways to establish a connection between the KMOC formula and the partial wave expansion. 
In our first approach, we begin with projecting out the components 
of $\vec{K}$ in \eqref{KMOC-full-CM-copy}. 
For each component, we use suitable Clebsch-Gordan relations for the base amplitudes and use a completeness relation 
involving the $\hat{k}'$-integral 
to compress the double sum (from $S$ and $S^\dagger$) 
to a single sum. Then we take the classical limit. The limit turns the sum into an integral. The spinor products are approximated by a Bessel function. 
As is well known (see, e.g., \cite{FORD1959,Berry:1972na}), the partial wave phase  in this limit converges to the radial action of the classical mechanics. 
We use this fact to confirm that \eqref{KMOC-full-CM-copy} agrees perfectly with the impulse computed from classical mechanics.

Our second approach is a slightly novel adaptation of the traditional saddle point approximation in the semi-classical limit \cite{FORD1959,Berry:1972na}. 
As explained in the original paper \cite{Kosower:2018adc}, the KMOC formula \eqref{KMOC-full-CM-copy} exhibits an ``interference" between $S$ and $S\dagger$ with $|\hbar\vec{q}| \ll |\vec{k}|$. 
In the classical limit, we can turn the displacement into a phase factor of the form $e^{i\vec{q}\cdot\vec{X}}$. 
Then the $q$-integral produces delta-functions, which forces the $\hat{k}'$-integral in \eqref{KMOC-full-CM-copy} to be localized in the direction dictated by the classical trajectory. 

Our two approaches lead to the same classical impulse as they should, but they give rather distinct perspectives. In the first one, the ``Clebsch-Gordan and completeness" process executes the interference between $S$ and $S^\dagger$. The remaining single sum is no longer rapidly oscillating, so there is no need for the saddle point approximation. 
In the second one, we tame the rapidly oscillating sums for $S$ and $S^\dagger$ separately by applying the saddle point approximation to fix their magnitudes and phases in the classical limit.

We put emphasis more on the general structure than on specific models. 
Our results are applicable to any elastic scattering process of two massive particles that can be treated in a post-Minkowskian framework. 
With some mild assumptions on the classical limit of the partial wave phases, all our arguments are non-perturbative in any coupling constant. 
To illustrate the ideas, we revisit two famous models 
that are exactly solvable in the probe (test-particle) limit: Coulomb scattering and charge-monopole scattering \cite{Schwinger:1976fr,Boulware:1976tv,Kol:2021jjc}. 

The charge-monopole scattering, to be discussed in section~\ref{sec:charge-monopole},  differs from the scalar scattering 
due to the angular momentum associated with the electromagnetic field. 
In the purely scalar case, the classical trajectory is confined on a plane, so the elastic scattering is completely solved by specifying the 
relation between the scattering angle $\theta$ and the impact parameter $b$. In contrast, for the charge-monopole scattering, there are two angles to be determined. Both of our methods successfully reproduce the two angles without any extra input from the geometry of the classical trajectory.

Our spinor basis, which builds upon \cite{Jiang:2020rwz,Csaki:2020inw}, is fully compatible with the Jacob-Wick helicity basis \cite{Jacob:1959at}, so it can express succinctly the partial wave amplitudes for particles with arbitrary spin. 
Although this paper only addresses scalar scattering and charge-monopole scattering, it is clear that our methods are applicable to gravitational spinning binaries. We hope to address such systems in a subsequent work. 

We conclude this introduction with a summary of our conventions. 
We use the mostly plus metric. Following \cite{Kosower:2018adc}, 
we denote the ``$(2\pi)$-normalized" quantities by hats, 
\begin{align}
    \hat{d}k = \frac{dk}{2\pi}\,,
    \quad 
    \hat{\delta}(k) = 2\pi \delta(k)\,, 
    \quad 
    \mbox{etc.} 
\end{align}
For a 3d vector $\vec{k}$, $\hat{k}$ denotes the unit vector $\vec{k}/|\hat{k}|$. 
The two unrelated uses of hats hopefully will not cause any confusion. 

\section{Classical impulse in the CM frame} \label{sec:KMOC-review}

\subsection{Kinematics of 2-to-2 elastic scattering} \label{sec:kinematics-22}

The Lorentz-invariant phase space measure of the 2-massive-particle Hilbert space is 
\begin{align}
d\mathrm{LIPS}_{1,2} = \widetilde{dk_1} \,\widetilde{dk_2} \,,
\quad 
\widetilde{dk_i} = \hat{d}^4 k_i \, \theta(k_i^0)\, \hat{\delta}(k_i^2+m_i^2) 
\quad (i=1,2) \,.
\end{align}
A covariant way to separate the CM degrees of freedom uses the change of variables, 
\begin{align}
     k_1 = \frac{1}{2}\left( 1 + \frac{m_1^2-m_2^2}{s} \right) P  + k \,,
    \quad
    k_2 &= \frac{1}{2}\left( 1 + \frac{m_2^2-m_1^2}{s} \right) P - k \,, 
    \quad s\equiv -P^2 > 0 \,.
\end{align}
It follows that 
\begin{align} \label{LIPS-CM_cov}
\begin{split}
        d\mathrm{LIPS}_{1,2}
    &= \hat{d}^4 P \hat{d}^4 k \, \theta(P^0) \,\hat{\delta}(2P\cdot k) \hat{\delta}(k^2 - \lambda(s,m_1^2,m_2^2)/4s) \,, 
    \\
    \lambda(x,y,z) &= x^2 + y^2 + z^2 - 2(xy +yz+zx) \,.
\end{split}
\end{align}
The appearance of $\lambda(x,y,z)$ should not come as a surprise because 
\begin{align}
    \sqrt{s}  = \sqrt{k^2 + m_1^2} + \sqrt{k^2 + m_2^2} 
    \quad \Longleftrightarrow \quad 
    k^2 = \lambda(s,m_1^2,m_2^2)/4s \,.
    \label{k-vs-s}
\end{align}
In \eqref{LIPS-CM_cov}, the integral measure is converted with unit Jacobian:
\begin{align}
\begin{split}
     dk_1 &= \frac{1}{2}\left( 1 + \frac{m_1^2-m_2^2}{s} \right) dP - \left(\frac{m_1^2-m_2^2}{2s^2}\right) ds P  + dk \,,
     \\
     dk_2 &= \frac{1}{2}\left( 1 + \frac{m_2^2-m_1^2}{s} \right) dP - \left( \frac{m_2^2-m_1^2}{2s^2} \right) ds P  - dk \,,
     \\
     & \quad \Longrightarrow \quad \hat{d}^4 k_1 \hat{d}^4 k_2 = \hat{d}^4 P \hat{d}^4 k \,.
\end{split}
\end{align}
The delta functions are also mapped with unit Jacobian since 
\begin{align}
    2P\cdot k = k_1^2 + m_1^2 - k_2^2 - m_2^2 \,, \quad k^2 - \frac{\lambda(s,m_1^2,m_2^2)}{4s} = \frac{1}{2} (k_1^2 + m_1^2 + k_2^2 + m_2^2) \,.
\end{align}
Finally, pulling out the CM part from \eqref{LIPS-CM_cov}, and setting $P^\mu|_{\mathrm{CM}} = (\sqrt{s},\vec{0})$, 
we obtain
\begin{align}
     d\mathrm{LIPS}_{1,2}|_{\mathrm{CM}} \equiv \hat{d}^4 k \, \hat{\delta}(2P\cdot k)_{\mathrm{CM}} \hat{\delta}(k^2 - \lambda(s,m_1^2,m_2^2)/4s) = \frac{|\vec{k}|}{4 \sqrt{s}}\, \hat{d} \Omega_{\mathrm{CM}} \,, 
     \label{LIPS-CM} 
\end{align}
where $|\vec{k}|$ and $\sqrt{s}$ are mutually dependent as in \eqref{k-vs-s}. 

\paragraph{Normalizing an operator} 

Consider any operator $X$ commuting with the translation generators $\mathbb{P}$. The matrix elements of $X$ can be written as 
\begin{align}
     \langle k_3, k_4 | X | k_1, k_2 \rangle 
     = \mathcal{N} \hat{\delta}^4(P_{34} - P_{12}) X_{34,12} \,.
     \label{N-def}
\end{align}
Here, $\mathcal{N}$ is a normalization factor to be determined shortly. 
In 4d, the operator $X$ and the matrix elements $X_{34,12}$ have the same mass dimension if and only if $\mathcal{N}$ is dimensionless. 

The simplest example is the identity operator: 
\begin{align}
\begin{split}
     \langle 3,4 | \mathbb{I} | 1,2 \rangle 
     &= \mathcal{N}\hat{\delta}^4(P_{34} - P_{12}) \mathbb{I}_{34,12}
     \\
     = \langle 3| 1\rangle  \langle 4| 2\rangle 
     &= 4 E_1 E_2 \hat{\delta}^3(\vec{k}_3 - \vec{k}_1) \hat{\delta}^3(\vec{k}_4 - \vec{k}_2) \,.
\end{split}
\end{align}
In the CM frame, \eqref{LIPS-CM_cov} and 
\eqref{LIPS-CM} implies that 
\begin{align}
    \mathcal{N} \, \mathbb{I}_{34,12} = \frac{4\sqrt{s}}{|\vec{k}|} 
    \hat{\delta}^2(\hat{k}_{34} - \hat{k}_{12})_\mathrm{CM} \,.
\end{align}
We find it convenient to fix $\mathcal{N}$ such that $\mathbb{I}_{34,12}$ takes the simplest form:
\begin{align}
   \mathcal{N} = \frac{4\sqrt{s}}{|\vec{k}|} 
   \quad \Longrightarrow \quad 
   \mathbb{I}_{34,12} = 
    \hat{\delta}^2(\hat{k}_{34} - \hat{k}_{12})_\mathrm{CM} \,. 
    \label{N-convention}
\end{align}

\paragraph{Unitarity for S} 
Many textbooks tend to switch too quickly from $S$ to $T$ via $S=1+iT$. 
We prefer working with $S$. 
If we follow the normalization convention in \eqref{N-def} and \eqref{N-convention}, 
\begin{align}
    \langle k_3, k_4 | S | k_1, k_2 \rangle = \frac{4\sqrt{s}}{|\vec{k}|} \,
 S_{34,12} \, \hat{\delta}^4(P_{34} - P_{12})  \,, 
\end{align}
the statement for the unitarity gives
\begin{align}
\begin{split}
   & \int  \widetilde{dk_3} \widetilde{dk_4}  \langle 1', 2' | S^\dagger | 3,4 \rangle \langle 3,4 | S |1,2 \rangle 
   = \langle 1' | 1 \rangle \langle 2' | 2 \rangle
   \\
   &\quad \Longrightarrow \quad 
   \int \hat{d}\Omega_{3,4} \, S^\dagger_{1'2',34} S_{34,12} = \hat{\delta}^2(\hat{k}_{1'2'} - \hat{k}_{12}) \,.
\end{split}
   \label{unitarity-S-CM}
\end{align}
In the CM frame, $S_{34,12}$ is a function of $\hat{k}_{34}$, $\hat{k}_{12}$, and $\sqrt{s}$.

\paragraph{Post-Minkowskian quantum mechanics}

In perturbative gravity, the dynamics of two massive bodies 
exchanging gravitons can be described in a post-Minkowskian framework 
with a Hamiltonian of the form 
\begin{align}
    H = \sqrt{\vec{k}^2 + m_1^2} + \sqrt{\vec{k}^2 + m_2^2} 
    - V(\vec{k},\vec{r}) \,.
    \label{PM-hamiltonian}
\end{align}
Here, $\vec{k}$ is the CM momentum and $\vec{r}$ is the relative coordinate between the two massive bodies canonically conjugate to $\vec{k}$. 
In this paper, we assume that a quantum 2-body system can be described by a Hamiltonian of this form \eqref{PM-hamiltonian} and try to extract classical observables from the quantum amplitudes.

\paragraph{Asymptotic in and out states}

In the standard non-relativistic scattering theory 
\cite{Taylor:1972pty}, the S-matrix is constructed from asymptotic in$(+)$ and out$(-)$ states, $|\vec{k}\pm\rangle = \Omega_\pm|\vec{k}\rangle$, 
which converge to the plane wave $e^{i\vec{k}\cdot\vec{r}}$ in the far past/future, respectively.

The normalized in-states should satisfy the orthogonality and completeness relations: 
\begin{align}
\begin{split}
     \int d^3r \, \psi_\mathrm{in}(\vec{k}_1;\vec{r})^* \psi_\mathrm{in}(\vec{k}_2;\vec{r}) &= \hat{\delta}^3(\vec{k}_1 - \vec{k}_2) \,,
     \\
     \int \hat{d}^3k \, 
     \psi_\mathrm{in}(\vec{k};\vec{r}_1) \psi_\mathrm{in}(\vec{k};\vec{r}_2)^* &= \delta^3(\vec{r}_1 -\vec{r}_2) \,.
\end{split}
\label{ortho-compl-in}
\end{align}
For scalar wave-functions, the out-states can be constructed from the in-states:
\begin{align}
    \psi_\mathrm{out}(\vec{k},\vec{r}) = \left[ \psi_\mathrm{in}(-\vec{k},\vec{r}) \right]^* \,.
    \label{out-from-in}
\end{align}
It follows that the out-states also satisfy the orthogonality-completeness relations \eqref{ortho-compl-in}. 

When the particles carry non-zero spin, 
the wave-functions should form representations of the $SU(2)$ spin group of the CM frame. 
We will not discuss spinning particles in detail in this paper. 

\paragraph{S-matrix from asymptotic states} 

Equipped with the in- and out-states, 
we can compute the S-matrix elements in the 3d sense by an overlap integral:
\begin{align}
    \mathcal{S}(\vec{k}', \vec{k}) = \int d^3\vec{r} \, \psi_\mathrm{out}(\vec{k}';\vec{r})^* \psi_\mathrm{in}(\vec{k};\vec{r}) \,.
    \label{S-as-overlap}
\end{align}
The orthogonality-completeness relations of the in/out-states guarantee unitarity:
\begin{align}
 \int \hat{d}^3 k' \,   \mathcal{S^\dagger}(\vec{k}'', \vec{k}') \mathcal{S}(\vec{k}', \vec{k}) = \hat{\delta}^3(\vec{k}'' - \vec{k})\,.
\end{align}
To make contact with the CM kinematics discussed earlier, we separate, in the $\vec{k}$-space, the radial dependence from 
the angular dependence . 
\begin{align}
   \mathcal{S}(\vec{k}', \vec{k}) = \frac{1}{|\vec{k}|^2} \hat{\delta}(|\vec{k}'| - |\vec{k}|) \, S(\vec{k}', \vec{k}) \,. 
   \label{SS-vs-S}
\end{align}
In terms of $S$, unitarity is stated just as in \eqref{unitarity-S-CM}, 
\begin{align}
\begin{split}
   \int \hat{d}\Omega' \, S^\dagger(\vec{k}'', \vec{k}') S(\vec{k}', \vec{k}) = \hat{\delta}^2(\hat{k}'' - \hat{k}) \,. 
\end{split}
   \label{unitarity-S-NR}
\end{align}

\paragraph{Probe limit}
The notion of ``quantum mechanics in the CM frame" from a 2-body system is quite general, but for simplicity of computation, our concrete examples will 
rely on the probe limit (a.k.a. test-particle limit or no-recoil limit). 
In view of the general form of the post-Minkowskian Hamiltonian \eqref{PM-hamiltonian}, the probe Hamiltonian can be obtained from the limit, 
\begin{align}
    H_{\mathrm{probe}} = \sqrt{\vec{k}^2 +m_1^2} - \lim_{m_2 \rightarrow \infty} V(\vec{k},\vec{r}) \,.
\end{align}

\subsection{KMOC in the CM frame}

A major goal of this paper is to establish 
a connection between the KMOC formula for the impulse 
and the partial wave expansion. To do so, we should first review the  formula. 
The key idea of the KMOC formula can be schematically written as
\begin{align}
    \Delta p_1 = \langle \psi | S^\dagger \mathbb{P}_1 S | \psi \rangle -  \langle \psi | \mathbb{P}_1 | \psi \rangle \,.
    \label{KMO-SS}
\end{align}
Here, $|\psi\rangle$ represents the initial state containing information on the two momenta ($p_1$, $p_2$) 
and the impact parameter vector $b$ subject to $p_1\cdot b = 0 = p_2 \cdot b$. 

The derivation of the KMOC formula includes a discussion on the wave-packets for the particles and their classical limit (more on this below). After the classical limit is taken, the wave-packets disappear and leave a concise formula:
\begin{align}
     (\Delta p_1)^\mu  = \frac{\hbar^2}{4} \int \hat{d}^4 q \, \hat{\delta}(p_1\cdot q) \hat{\delta}(p_2 \cdot q) \, e^{iq\cdot b} K^\mu(p_1, p_2 ; q) \,,
\end{align}
where $K^\mu(p_1, p_2 ; q)$ is defined in two steps: 
\begin{align}
\begin{split}
     &\int \widetilde{dp_3}\widetilde{dp_4} \langle 1', 2'| S^\dagger | 3,4 \rangle  (p_3 - p_1)^\mu  
     \langle 3,4  | S | 1,2 \rangle =  \widetilde{K}^\mu(1',2';1,2) \, \hat{\delta}^4(P_{1'2'} - P_{12})   
     \\
     &\Longrightarrow \quad K^\mu(p_1, p_2 ; q) = \widetilde{K}^\mu(p_1 + \hbar q/2, p_2-\hbar q/2 ; p_1 -\hbar q/2, p_2 + \hbar q/2) \,.
\end{split}
\end{align}
\paragraph{CM frame} 
So far, we have maintained manifest Lorentz covariance. We can specialize to the CM frame using the notations developed in section~\ref{sec:kinematics-22}. 
The key results can be summarized as 
\begin{align}
\begin{split}
         \Delta \vec{k}  &= \frac{\hbar^2}{|\vec{k}|^2} \int \hat{d}^2 q \,e^{i \vec{q}\cdot \vec{b} } \,\vec{K} (\vec{k},\vec{q}) \,,
         \\
         \vec{K}(\vec{k},\vec{q})  &= \int \hat{d}^2 \hat{k}_2 \left[ S^\dagger(\vec{k}_3,\vec{k}_2) (\vec{k}_2 - \vec{k}) S(\vec{k}_2, \vec{k}_1) \right] \,,
         \\
         \vec{k}_3 &= \vec{k} + \hbar \vec{q}/2\,, \quad \vec{k}_1 = \vec{k} - \hbar \vec{q}/2\,.
\end{split}
\label{KMOC-full-CM}
\end{align}
Here, $\vec{k}$ is the relative 3-momentum in the CM frame for (incoming) particles 1 and 2. It is understood that the wave-vector $\vec{q}$ is subject to the constraint $\vec{k}\cdot \vec{q} = 0$, just as the impact parameter $\vec{b}$ is. 
The formula \eqref{KMOC-full-CM} depends on the energy $(\sqrt{s})$ through the S-matrix.

\paragraph{Wave packet for classical particles}

In the KMOC formulation, the classical limit requires wave packets for the particles 
at the intermediate stage. Let us briefly review how it works. 
In the non-relativistic setup, we assume a Gaussian wave packet of the form 
\begin{align}
    \phi(\vec{k})_{\vec{b}} \equiv  \mathcal{N} e^{-(\vec{k}-\vec{k}_0)^2/2w^2 - i\vec{k}\cdot\vec{b}/\hbar} \,.
   \label{k-Gaussian}
\end{align}
The modulation $e^{-i\vec{k}\cdot\vec{b}/\hbar}$ accounts for the displacement by $\vec{b}$ in the $\vec{r}$-space. 

In the derivation of the KMOC impulse formula, we encounter the product 
\begin{align}
    \phi(\vec{k} + \hbar\vec{q}/2  )^* \,\phi(\vec{k} - \hbar\vec{q}/2 ) = |\mathcal{N}|^2 e^{-(\vec{k}-\vec{k}_0)^2/w^2-\vec{q}^2 \hbar^2 /4w^2 + i\vec{q}\cdot\vec{b} }  \,.
\end{align}
The key assumption that defines the classical limit is 
\begin{align}
 \frac{\hbar}{|\vec{b}|}  \sim \hbar |\vec{q}|  \ll   |\vec{k}_0| \,.
 \label{classical-hierarchy}
\end{align}
It allows us to choose the width parameter $w$ within the range, 
\begin{align}
   \frac{\hbar}{|\vec{b}|} \ll w \ll |\vec{k}_0| \,.
\end{align}
This choice effectively results in the replacement, 
\begin{align}
    |\mathcal{N}|^2 e^{-(\vec{k}-\vec{k}_0)^2/w^2 } \; \rightarrow \;\; \hat{\delta}^3(\vec{k}-\vec{k}_0) 
    \,,
    \quad 
     e^{-\vec{q}^2 \hbar^2 /4w^2 + i\vec{q}\cdot\vec{b} } \; \rightarrow \;\; e^{ i\vec{q}\cdot\vec{b} }  \,.
\end{align}
After the replacement, the integral reduces to 
\begin{align}
\begin{split}
     &\qquad \int \hat{d}^3 k_3 \hat{d}^3 k_1 \phi^*(k_3) \phi(k_1) F(\vec{k},\vec{q},\cdots) 
     \\
     &\rightarrow \;\;  
    \hbar^3 \int \hat{d}^3 k\, \hat{d}^3 q  \left[ \hat{\delta}^3(\vec{k}-\vec{k}_0) e^{ i\vec{q}\cdot\vec{b} } \right] F(\vec{k},\vec{q},\cdots) 
    =  \hbar^3 \int \hat{d}^3 q \, e^{ i\vec{q}\cdot\vec{b} } \, F(\vec{k}_0,\vec{q},\cdots) \,. 
\end{split}
\end{align}
Dropping the subscript from $\vec{k}_0$, we arrive at the formula \eqref{KMOC-full-CM}.

Another crucial implication of the hierarchy of the scales \eqref{classical-hierarchy} is that the angle between $\vec{k}_3$ and 
$\vec{k}_1$, defined as $\cos\alpha = \hat{k}_3 \cdot \hat{k}_1$, can be always assumed to be infinitesimal:  
\begin{align}
    \alpha = 2 \arctan\left(\frac{\hbar |\vec{q}|}{2|\vec{k}|}\right) 
    = \frac{\hbar |\vec{q}|}{|\vec{k}|} + \mathcal{O}\left(\frac{\hbar |\vec{q}|}{|\vec{k}|}\right)^3\,. 
    \label{small-alpha}
\end{align}
The sub-leading terms will not contribute to the classical impulse, 
as long as we focus strictly on the conservative sector.

\section{Partial wave expansion with spinors} \label{sec:partial-wave}

We review the partial wave expansion in quantum mechanics, with emphasis on a spinor basis for base amplitudes, 
following \cite{Csaki:2020inw,Jiang:2020rwz}. 
Our main interest lies in the 2-body quantum mechanics in the CM frame, but the spinor basis could be used in any quantum mechanics problem with rotation symmetry. 

\subsection{Helicity basis} \label{sec:helicity}

\paragraph{Spinors on the sphere}  

For a given point on the ``Bloch" sphere, 
\begin{align}
    \hat{k} = (\sin\theta\cos\phi, \sin\theta\sin\phi, \cos\theta)\,, 
\end{align}
we consider the matrix $(\hat{k}\cdot \vec{\sigma})$ and its decomposition in the eigen-spinor basis, 
\begin{align}
    \hat{k} \cdot \vec{\sigma}
    = \begin{pmatrix}
    \cos\theta & e^{-i\phi} \sin\theta  \\
    e^{+i\phi}\sin\theta  & -\cos\theta
    \end{pmatrix} = 
    | k^+ )( k_+ | - | k^- )( k_- | \,.
\end{align}
In labelling the spinors, we removed hats from the unit vectors, as no confusion is likely. 
The components of the spinors can be written explicitly as 
\begin{align}
\begin{split}
    \hat{k} \cdot \vec{\sigma}
    &= 
    \begin{pmatrix}
    c_{\theta/2} \\ e^{+i\phi} s_{\theta/2} 
    \end{pmatrix}
    \begin{pmatrix}
    c_{\theta/2} & e^{-i\phi} s_{\theta/2} 
    \end{pmatrix}
    - 
     \begin{pmatrix}
    - e^{-i\phi} s_{\theta/2} \\ c_{\theta/2}
    \end{pmatrix}
    \begin{pmatrix}
    - e^{+i\phi} s_{\theta/2} & c_{\theta/2}
    \end{pmatrix}
    \\
    &= 
    \begin{pmatrix}
    e^{-i\phi} c_{\theta/2} \\  s_{\theta/2} 
    \end{pmatrix}
    \begin{pmatrix}
    e^{+i\phi} c_{\theta/2} &  s_{\theta/2} 
    \end{pmatrix}
    - 
     \begin{pmatrix}
     - s_{\theta/2} \\ e^{+i\phi} c_{\theta/2}
    \end{pmatrix}
    \begin{pmatrix}
    - s_{\theta/2} &  e^{-i\phi} c_{\theta/2}
    \end{pmatrix}
    \,.
\end{split}
\label{n-dot-sigmax}
\end{align}
The shorthand notations are self-explanatory: $c_{\theta/2} = \cos(\theta/2)$, $s_{\theta/2} = \sin(\theta/2)$.

In \eqref{n-dot-sigmax} we wrote the same result in two different ways for a reason. 
In the first line, the components $c_{\theta/2}$ and $e^{\pm i\phi} s_{\theta/2}$ are non-singular in the northern hemisphere, 
while in the second line, the components $s_{\theta/2}$ and $e^{\pm i\phi} c_{\theta/2}$ are non-singular in the southern hemisphere. 

The transition between the two hemispheres show that, although the matrix $(\hat{k}\cdot \vec{\sigma})$ is globally well-defined, 
the spinors $|k^\pm )$ are only locally well-defined and should be regarded as sections of a bundle. 
In a physics language, we may say the spinors are defined up to a ``gauge ambiguity".

Relations among spinors and their products must be covariant under the gauge ambiguity; 
the LHS and RHS of an equality should transform in the same way. 
Some products of the spinors are manifestly free of the gauge ambiguity. Examples include
\begin{align}
    |k^+ )\otimes |k^- ) \,,
    \quad 
    |k^+ ) ( k_+ | \,,
    \quad 
    |k^- ) ( k_- | \,.
\end{align}

When we deal with only two unit vectors at a time, we find it convenient to choose the coordinate axes such that 
both unit vectors lie on the $(x,z)$-plane. We call this choice ``co-planar gauge". 
The co-planar gauge can be extended to more than two unit vectors, provided that they lie on the same great circle on the sphere.

\paragraph{Tensor product and total symmetrization} 

For an arbitrary choice of $\hat{k}$, the spinors $|k^\pm )$ form an orthonormal basis of the spin-$1/2$ representation. 
The key idea behind the spinor technology is simply that 
it is convenient to build up ``spin $n$" representations from the tensor product of $(2n)$ copies 
of the spin-$(1/2)$ representation:
\begin{align}
    (1/2)^{\otimes 2n} = n \oplus \mbox{(smaller reps)}. 
\end{align}
To separate the spin-$n$ representation from the rest, we simply project onto the totally symmetric subspace. 

To illustrate the idea, let us use the standard basis of the $n=1/2$ rep,
\begin{align}
    |+) = \begin{pmatrix}
        1 \\ 0 
    \end{pmatrix} \,,
    \quad 
    |-) = \begin{pmatrix}
        0 \\ 1 
    \end{pmatrix} \,,
\end{align}
as building blocks to construct the basis of higher reps. 
We propose 
\begin{align}
    |n, a ) = \left[ |+)^{n+a} |-)^{n-a} \right]_\mathrm{ts} \,, 
\end{align}
where ``ts" stands for the unit-normalized totally symmetric tensor product defined by 
\begin{align}
   \left[ |+)^{n+a} |-)^{n-a} \right]_\mathrm{ts} = \binom{2n}{a}^{-1/2} \sum_{\mathcal{S} \in \mathrm{DP}} 
   \mathcal{S} \left[ |+)^{\otimes(n+a)} \otimes |-)^{\otimes(n-a)} \right]\,.
   \label{ts2}
\end{align}
Here, the permutation $\mathcal{S}$ runs over distinct permutations (DP) only, and all terms in the sum carry unit weight. 
For example, 
\begin{align}
    \sum_{\mathcal{S} \in \mathrm{DP}}   \mathcal{S} \left[ |+)^{\otimes 2} \otimes |-)\right] = 
    |++- ) + |+-+ ) + |-++ ) \,, 
\end{align}
where the tensor product notation is suppressed on the RHS. 
The normalization factor in \eqref{ts2} simply reflects the number of distinct permutations. 

Next, we consider an SU(2) rotation $U$ taking the basis vectors from the $\hat{z}$ axis to another axis $\hat{k}$ along the geodesic. 
A key point is that the total symmetrization and the rotation commute such that 
\begin{align}
\begin{split}
        R_{n\mathrm{-rep}} |n,a ) &= (U^{\otimes 2n})_\mathrm{ts} \left[ |+)^{n+a} |-)^{n-a} \right]_\mathrm{ts}
        \\
        &= \left[ U|+)^{n+a}  U |-)^{n-a} \right]_\mathrm{ts}
        = \left[ |k^+)^{n+a} |k^-)^{n-a} \right]_\mathrm{ts} \,.
\end{split}
\end{align}

In the partial wave analysis, which we will discuss shortly, 
typical textbook treatments begin with ``momentum direction" basis $|\hat{k}\rangle$ 
and switch to angular momentum basis $|n,m)$ via spherical harmonics. 
We will argue that a more efficient alternative is to 
replace spherical harmonics by products of spinors.

\paragraph{Spinor basis} 

Let us introduce a shorthand notation for the ``spin $n$" spinor basis 
and their inner product, 
\begin{align}
 |k \r)^n_a = \left[ |k^+)^{n +a} |k^- )^{n-a} \right]_\mathrm{ts} \,, 
\quad 
    \l(k_2 |k_1\r)^n_{b,a}  = \l(k_2|^n_b |k_1\r)^n_a \,.
    \label{spinor-prod}
\end{align}
Here and below, $n, a, b \in \textstyle{\frac{1}{2}}\mathbb{Z}$, while  
$n-a, n-b \in \mathbb{Z}$. 
The simplest example of this spinor product is for $a=b=0$: 
\begin{align}
    P_{n}(\cos\theta) = 
    \l(k_2 | k_1 \r)^n_{0,0} \,, 
    \quad 
    \hat{k}_2 \cdot \hat{k}_1 = \cos\theta \,.
    \label{Legendre-k}
\end{align} 
The spinor products will serve as the base amplitudes of the partial wave expansion. 

From the fact that
\begin{align}
    \begin{pmatrix}
        (k_{2+}| k_{1}^+) & (k_{2+}|k_{1}^-) 
        \\
        (k_{2-}| k_{1}^+) & (k_{2-}|k_{1}^-) 
    \end{pmatrix}
    \in 
    \mathrm{SU}(2) \,,
\end{align}
it follows that 
\begin{align}
   (k_{2+}|k_{1}^+) = (k_{2-}|k_{1}^-)^*\,, \quad 
   (k_{2+}|k_{1}^-) = - (k_{2-}|k_{1}^+)^* 
   \,,
   \label{spinor-cc}
\end{align}
which in turn implies that 
\begin{align}
    \l(k_2 |k_1\r)^n_{b,a}  = (-1)^{b-a} \l(k_1 |k_2\r)^n_{-a,-b}  \,.
    \label{spinor-prod-reflection}
\end{align}

\paragraph{Wigner and Jacobi} 

We first recall the definitions of the Wigner D and d matrices:
\begin{align}
\begin{split}
    D^n_{b,a}(\alpha,\beta,\gamma) &= (n,b| \mathcal{R}(\alpha, \beta, \gamma) |n,a)  
    = e^{-ib\alpha} d^n_{b,a}(\beta) e^{-ia\gamma} \,,
    \\
    \mathcal{R}(\alpha, \beta, \gamma) &= e^{-i\alpha J_z} e^{-i\beta J_y} e^{-i\gamma J_z} \,. 
\end{split}
\end{align}
It is clear that our spinor products \eqref{spinor-prod} are 
physically equivalent to Wigner D matrices upon a suitable mapping of angles. 
We find it convenient to discuss the d-functions $d^n_{b,a}(\beta)$ 
separately from the phase factors $e^{-ib \alpha}$, $e^{-ia\gamma}$.

Suppose we align the two vectors such that 
\begin{align}
\begin{pmatrix}
        (k_{2+}| k_{1}^+) & (k_{2+}|k_{1}^-) 
        \\
        (k_{2-}| k_{1}^+) & (k_{2-}|k_{1}^-) 
    \end{pmatrix}
    =
    \begin{pmatrix}
        c_{\theta/2} & -s_{\theta/2}
        \\
        s_{\theta/2} & c_{\theta/2}
    \end{pmatrix} 
    = 
        \begin{pmatrix}
        d^{1/2}_{1/2,1/2}(\theta) & d^{1/2}_{1/2,-1/2}(\theta) \\
        d^{1/2}_{-1/2,1/2}(\theta) & d^{1/2}_{-1/2,-1/2}(\theta)
    \end{pmatrix} \,. 
\end{align}
The total symmetrization construction suggests that
$\l(k_2 | k_1 \r)^n_{b,a} = d^n_{b,a}(\theta)$ for all $n$, $a$, $b$. 
An explicit proof of this equality and related facts are given in appendix~\ref{app:wave-ftn}. 

In general, the spinor product takes the form
\begin{align}
    \l(k_2 | k_1 \r)^n_{b,a} = d^n_{b,a}(\theta) e^{i\chi} 
    \,,
    \quad 
    \hat{k}_2 \cdot \hat{k}_1 = \cos\theta\,. 
\end{align}
The d-function determines the absolute value of the spinor product, but we should also identify the phase factor. The phase $\chi$ depends not only on the vectors $\hat{k}_{1,2}$ but also on some auxiliary choices. 
As we already mentioned, there is a ``gauge ambiguity" between coordinate patches in a fixed coordinate frame. We may also rotate the coordinate axes slightly while keeping $\hat{k}_{2,1}$ fixed. 

Of course, these auxiliary choices should not affect the physical observables. 
In a typical textbook discussion of spinor spherical harmonics 
and partial wave expansion, 
these ambiguities are ``gauge-fixed" in some way. 
We advocate the viewpoint that it is better to leave the ambiguities unfixed. 
As long as the two sides of an equality have the same ambiguity, 
the equality is valid. 
We will see below that ``living with gauge ambiguity" simplifies 
and clarifies many steps in computations.

\paragraph{Orthogonality and Completeness for spinors} 

The Wigner D-matrices are known to satisfy orthogonality and completeness relations. Let us state the relations in our spinor notation. 
The orthogonality can be written as
\begin{align}
  \pi \int \hat{d}\Omega_k |k\r)^m_a  \l(k|^n_a =  \frac{\delta_{m, n}}{2n+1}  \, I_{(2n+1)\times (2n+1)} \,.
   \label{ortho-spinor}
\end{align}
The fact that the LHS is proportional to $\delta_{m,n} I_{(2n+1)\times (2n+1)} $ is a direct consequence of Schur's lemma. 
The normalization is then easily fixed by taking the trace of both sides. 

The completeness relation can be written as 
\begin{align}
    \pi \sum_{n=|a|}^\infty (2n+1) 
   \l(k_2|k_1\r)^n_{a,a}
    = \hat{\delta}^2(\hat{k}_2 - \hat{k}_1)  \,.
    \label{completeness-spinor}
\end{align}
Here is the sketch of a proof of \eqref{completeness-spinor}.
The LHS is an SU$(2)$ scalar up to the frame dependence we discussed above. 
So, up to an {\em overall} phase, we can take it to the co-planar gauge, 
where we recognize each term in the sum as a Jacobi polynomial. 
Then, Sturm-Liouville theory (Christoffel-Darboux formula in particular) applied to Jacobi polynomials offers a rigorous proof of the completeness. The overall phase is irrelevant when $\hat{k}_2 \neq \hat{k}_1$ and it vanishes when $\hat{k}_2 = \hat{k}_1$, so the proof is gauge-independent.

\subsection{Partial wave in the helicity basis}

We explain how to rewrite the partial wave expansion in terms of the spinor products. 
We begin with the scalar case and proceed to processes with spin. 

\paragraph{Partial wave sum for asymptotic states}

The partial wave expansion begins with an identity valid for a free particle wavefunction:
\begin{align}
    e^{i\vec{k}\cdot\vec{r}} = \sum_{n=0}^\infty (2n+1) i^n j_n(kr) P_n(\hat{k}\cdot \hat{r}) \,.
\end{align}
In an interacting theory with spherical symmetry, we construct the asymptotic in and out states, which must approach the plane-wave as $\vec{k}\cdot \vec{r} \rightarrow -\infty$ (in) or $\vec{k}\cdot \vec{r} \rightarrow +\infty$ (out). 

Recall the asymptotic formula for the spherical Bessel function, 
\begin{align}
    j_n(x) \approx \frac{1}{x} \sin\left( x- \frac{\pi n}{2} \right)  \,, 
    \quad x \rightarrow  +\infty \,.
    \label{sph-bessel-asymp}
\end{align}
We normalize the interacting radial wave-function $R_n(kr)$ such that it is real-valued for all $r$ and 
that it is approximated by 
\begin{align}
    R_n(x) \approx \frac{1}{x} \sin\left( x + \delta_n - \frac{\pi n}{2} \right)  \,, 
    \quad x \rightarrow +\infty \,.
    \label{delta-n-def}
\end{align}
With this normalization, we may use the spinor product \eqref{spinor-prod} or \eqref{Legendre-k} to write the asymptotic incoming wave as 
\begin{align}
    \psi_\mathrm{in}(\vec{k} ;\vec{r}) =  \sum_{n=0}^\infty (2n+1) i^n  e^{i\delta_n} R_n(kr) \l( r|k \r)^n_{0,0} \,.
\label{in-partial}
\end{align}

The wave-function \eqref{in-partial} is incoming in the sense that, for $kr\rightarrow \infty$, 
\begin{align}
    e^{i\delta_n} R_n(kr) \sim \frac{1}{2ikr} \left( -
    e^{-i(kr-\pi n/2)} + e^{2i\delta_n } e^{+i (kr-\pi n/2)} \right) \,.
\end{align}
Similarly, we can write the outgoing wave as
\begin{align}
     \psi_\mathrm{out}(\vec{k} ;\vec{r}) = \sum_{n=0}^\infty (2n+1)  i^n e^{-i\delta_n} R_n(kr) \l( r|k \r)^n_{0,0} \,.
     \label{out-partial}
\end{align}
Using some elementary properties of spinor products, 
\begin{align}
  |\mathrm{-}k \r)^n_0 = (-1)^n |k \r)^n_0 \,,
  \quad 
\l( r|k \r)^n_{0,0} = P_n(\hat{k}\cdot\hat{r}) = \l( k|r \r)^n_{0,0} \,,
\end{align} 
we can verify the general relation \eqref{out-from-in} at the partial wave level: 
\begin{align}
\begin{split}
    \psi_\mathrm{in}(-\vec{k} ;\vec{r}) &=  \sum_{n=0}^\infty (2n+1) i^n  e^{i\delta_n} R_n(kr) \l( r|\mathrm{-}k \r)^n_{0,0} 
    \\
    &=  \sum_{n=0}^\infty (2n+1) (-i)^n  e^{i\delta_n} R_n(kr) \l( k|r \r)^n_{0,0} 
    = \left[ \psi_\mathrm{out}(\vec{k} ;\vec{r}) \right]^* 
    \,.
\end{split}
\end{align}

Having prepared the asymptotic states, we can use \eqref{S-as-overlap} and \eqref{SS-vs-S} to compute the S-matrix. 
Using the orthogonality \eqref{ortho-spinor} adapted to the $\vec{r}$-space and the normalization 
of the radial wave-function in the form 
\begin{align}
    \int_0^\infty r^2 dr \, R_n(k'r) R_n(kr) = \frac{1}{4k^2} \hat{\delta}(k'-k) \,,
    \label{radial-ortho}
\end{align}
we reach the final expression for the partial wave expansion of the S-matrix, 
\begin{align}
   S(\vec{k}', \vec{k}) = \pi \sum_{n=0}^\infty (2n+1) e^{2i\delta_n} \l( k'| k \r)^n_{0,0} \,.
   \label{S-partial-scalar}
\end{align}

\paragraph{Particles with spin: Jacob-Wick helicity basis} 

To generalize the partial wave expansion to include the spins of the particles, 
we will take the spinor product \eqref{spinor-prod} as the ``base amplitude" 
of the Jacob-Wick helicity basis \cite{Jacob:1959at}. 

In the CM frame, the spatial momenta of the two incoming particles are co-linear. The same holds for the outgoing particles. It makes sense to 
enumerate the spin states of the particles according to the helicity with respect to the momentum. 
Let $s_1$, $s_2$ be the spins of the two particles. For elastic scattering, 
it suffices to discuss the incoming particles. The helicity labels $a_{1,2}$  run in integer spacing within the range $a_i \le |s_i|$. 

\begin{figure}[ht]
\begin{center}
    \includegraphics[width=12.5cm]{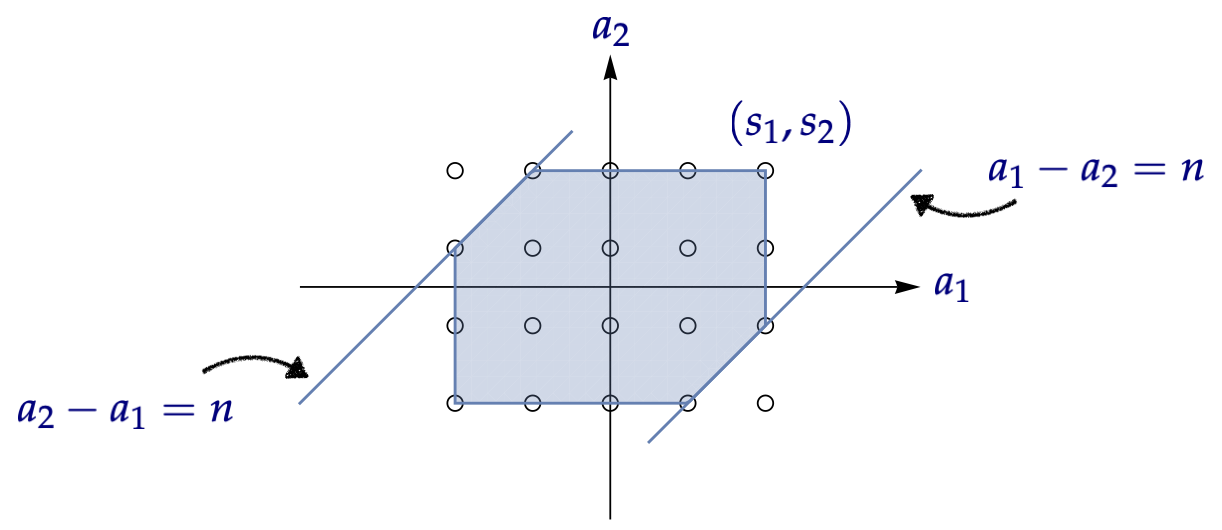}
\end{center}
\vskip -1cm
\caption{The reduction of the size of the matrix $M$. }
\label{counting-JW}
\end{figure}

In these notations, the S-matrix in the Jacob-Wick basis takes the form 
\begin{align}
\begin{split}
     S(\vec{k}', \vec{k})^{a_3a_4}{}_{a_1a_2} = \pi \sum_{n=n_*}^\infty (2n+1) (M_n)^{a_3a_4}{}_{a_1a_2} \l(k' |k \r)^n_{a_3-a_4,a_1-a_2} \,.
\end{split}
\label{S-partial-spin}
\end{align}
For fixed helicity labels, the sum over $n$ begins at $n_* = \max(|a_3-a_4|,|a_1-a_2|)$. 
The size of the matrix $M_n$, denoted by $d(M_n)$, depends on $n$. For $n \ge s_1 +s_2$, it takes the generic value $(2s_1+1)(2s_2+1)$. 
For smaller $n$, the value is smaller. 
In general, $d(M_n)$ is determined by the two conditions, 
\begin{align}
    |a_i|\le s_i \,,
    \quad 
    n \ge |a_1 - a_2| \,.
\end{align}
With the notations, $s_+ = s_1 + s_2$, $s_- = |s_1 -s_2|$, $s_\mathrm{m} = \min(s_1,s_2)$, 
we can write down a concise counting formula:
\begin{align} \label{dim-M}
d(M_n) =
\begin{cases}
(2n+1)(2s_\mathrm{m} +1), \hspace{2.7cm} 0 \le n \le s_- 
\\
(2s_{1}+1)(2s_{2}+1) - m(m+1), \quad s_- \le n \le s_+ \,,\quad m = s_+ - n \\
(2s_{1}+1)(2s_{2}+1), \hspace{2.7cm} s_+ \le n
\end{cases}
\end{align}
Pictorially, we are counting the lattice points in the shaded area (including the boundary) in Fig.~\ref{counting-JW}, which shows an example with $s_1=2$, $s_2=3/2$, $n=5/2$.

\paragraph{Partial wave unitarity} 

It is instructive to see how the unitarity, $S^\dagger S = I$, works at the partial wave level. The conjugate of the S-matrix \eqref{S-partial-spin} can be written as 
\begin{align}
\begin{split}
     S^\dagger(\vec{k}'', \vec{k}')^{a_1''a_2''}{}_{a_3a_4} = \pi \sum_{m=m_*}^\infty (2n'+1) (M_{m}^\dagger)^{a_1''a_2''}{}_{a_3a_4} \l(k'' |k' \r)^m_{a_1''-a_2'',a_3-a_4,} \,.
\end{split}
\label{S-partial-spin-dagger}
\end{align}
In computing $S^\dagger S$, the first step is to integrate over $\hat{k}'$. The orthogonality \eqref{ortho-spinor} gives 
\begin{align}
  \pi \int \hat{d}^2 k' |k'\r)^m_{a_3-a_4} \l(k'|^n_{a_3-a_4} =  \frac{\delta_{m, n}}{2n+1}  \, I_{(2n+1)\times (2n+1)} \,, 
   \label{unitarity-ortho}
\end{align}
leaving behind the product $\l(k''|k\r)^n_{a_1''-a_2'',a_1-a_2}$. 
The second step is to require the unitarity for $M$ (for elastic scattering) in the sense that
\begin{align}
    \sum_{a_3, a_4} (M_n^\dagger)^{a_1''a_2''}{}_{a_3a_4} (M_n)^{a_3a_4}{}_{a_1a_2} = \delta^{a_1''}{}_{a_1} \delta^{a_2''}{}_{a_2} \,.
    \label{unitarity-M}
\end{align}
The final step uses the completeness relation \eqref{completeness-spinor}, which yields  
\begin{align}
    \pi \sum_{n=|a_1 -a_2|}^\infty (2n+1) 
   \l(k''|k\r)^n_{a_1 -a_2,a_1 -a_2}
    = \hat{\delta}^2(\hat{k}'' - \hat{k})    \,.
    \label{unitarity-completeness}
\end{align}
The lower bound of the sum in \eqref{unitarity-completeness} 
is compatible with the reduction of the size of the matrix $M_n$ in \eqref{dim-M}.

In summary, we have shown that the partial wave unitary \eqref{unitarity-M}, 
taking the reduction \eqref{dim-M} into account,  
ensures the unitarity in the standard form,
\begin{align}
\begin{split}
   \sum_{a_3,a_4} \int \hat{d}\Omega' \, S^\dagger(\vec{k}'', \vec{k}')^{a_1'' a_2''}{}_{a_3 a_4} S(\vec{k}', \vec{k})^{a_3 a_4}{}_{a_1 a_2 } = \delta^{a_1''}{}_{a_1} \delta^{a_2''}{}_{a_2} \hat{\delta}^2(\hat{k}'' - \hat{k}) \,. 
\end{split}
   \label{unitarity-S-spin}
\end{align}

\paragraph{Charge-monopole scattering}

The Jacob-Wick helicity basis conveniently captures 
both the orbital angular momentum and the spins of the particles. 
It was shown in \cite{Csaki:2020inw} that the same basis can also capture 
the angular momentum produced by the electromagnetic fields 
of the charge-monopole system. For electric charge $e$ and magnetic charge $g$, the quantized angular momentum is $h=eg/4\pi \hbar$. 
To incorporate its effect in the partial wave expansion, one simply shifts the helicity of the incoming and outgoing particles by 
\begin{align}
\begin{split}
     S(\vec{k}', \vec{k})^{a_3a_4}{}_{a_1a_2} = \pi \sum_{n=n_*}^\infty (2n+1) (M_n)^{a_3a_4}{}_{a_1a_2} \l(k' |k \r)^n_{a_3-a_4-h,a_1-a_2+h} \,.
\end{split}
\label{S-partial-spin-h}
\end{align}
The reduction of the size of $M$ should be modified accordingly. 
In section~\ref{sec:charge-monopole}, we will study in detail a simple example where the charge and the monopole have both spin zero.

\section{Classical impulse from partial wave} \label{sec:KMOC-PW}

In the scalar case, the dynamics of the system is encoded in the 
partial wave phase $e^{2i\delta_n}$. 
The standard WKB argument shows that the large $n$ limit of $2\delta_n$ gives the radial action of the classical mechanics. 
In the amplitude context, the use of the radial action was advocated in \cite{Bern:2021dqo}.
Here, we review some key aspects of the radial action.

\subsection{Radial action}

When the Hamiltonian depends only on $\vec{r}^2$ and $\vec{p}^2$, 
the conservation of energy and angular momentum leads to a relation
\begin{align}
     p_r^2 +\frac{L^2}{r^2} = f(E,r) \,, 
\quad
    p_r = m \frac{dr}{dt} \,,
    \quad 
    L = mr^2 \frac{d\phi}{dt} \,.
         \label{pr-general}
\end{align} 
In Newtonian mechanics, we have
\begin{align}
    f(E,r) = 2m (E - V(r)) = k^2 -2m V(r) \,,
\end{align}
where $V(r)$ is the potential. 
The relation \eqref{pr-general} is equally valid in the post-Minkowskian setting with a suitable modification of $f(E,r)$. 
For example, for a Coulomb-like system, 
\begin{align}
    p_r^2 + \frac{L^2}{r^2} + m^2 = (\sqrt{k^2+m^2} - V(r))^2 \,.
    \label{Coulomb-like-RA}
\end{align}
Solving the relation \eqref{pr-general} for $p_r$ and using the dispersion relation $E(k)$, in what follows we regard $p_r$ as a function of $k, L$ and $r$. 

The radial action is a regularized round-trip 
integral of $p_r dr$ from the turning point (where $p_r=0$) to infinity:
\begin{align}
    I(k,L) = 2 \lim_{R\rightarrow \infty} \left( \int_{r_\mathrm{min}}^R p_r dr - k R \right)\,.
\end{align}
This expression is good for a fast-falling potential. For the Coulomb potential, $V(r)\sim r^{-1}$, we should further subtract a $\log(kR)$ term. 

The radial action is related to the scattering angle $\theta = \pi - \Delta\phi$ through 
\begin{align}
    d\phi = \frac{d\phi}{dr} dr = \frac{p_\phi}{p_r} dr
    \quad
 \Longrightarrow \quad \Delta\phi = 2 \int^\infty_{r_\mathrm{min}} d\phi = 2 \int^\infty_{r_\mathrm{min}} \frac{L/r^2}{p_r} dr = - \frac{\partial I}{\partial L} \,.
\label{Delta-phi-RA}
\end{align}
In other words, 
\begin{align}
    \theta = \theta_*(L) = \frac{\partial}{\partial L} \left[ \pi L + I(k,L) \right] = \frac{\partial}{\partial L}  \widehat{I}(L) \,.
    \label{theta-IL}
\end{align}
A Legendre transform then gives the inverse relation,  
\begin{align}
   Q(\theta) = \widehat{I}(L) - \theta L  = I(L) + (\pi - \theta ) L \quad \Longrightarrow \quad 
   L = L_*(\theta) = - \frac{\partial}{\partial\theta} Q(\theta) \,.
   \label{I-Q-Legendre}
\end{align}

\subsection{Coulomb scattering} 

As a concrete example of the scalar scattering, 
and to set a reference for the charge-monopole scattering 
in the next section, 
we review the relativistic Coulomb scattering, 
following \cite{Kol:2021jjc} closely.

\paragraph{Classical trajectory}

The classical equation of motion can be solved in the standard way. The scattering solution is given by, for example, \cite{boyer:2004}
\begin{align}
    \frac{1}{r} = \sqrt{\frac{E^2 L^2 -m^2 \left(L^2 -\alpha^2\right)}{\left(L^2 -\alpha^2\right)^2}} \cos \left[\sqrt{1-\left(\frac{\alpha}{L }\right)^2}\left(\phi-\phi_0\right)\right]+\frac{E \alpha}{L^2 -\alpha^2} \,, 
\end{align}
where $E$ is the conserved energy, $L$ is the conserved angular momentum, 
and $\alpha$ is the classical coupling constant.
The sign convention is such that $\alpha>0$ corresponds to an attractive potential. 

\begin{figure}[ht]
    \centering
    \includegraphics[width=10cm]{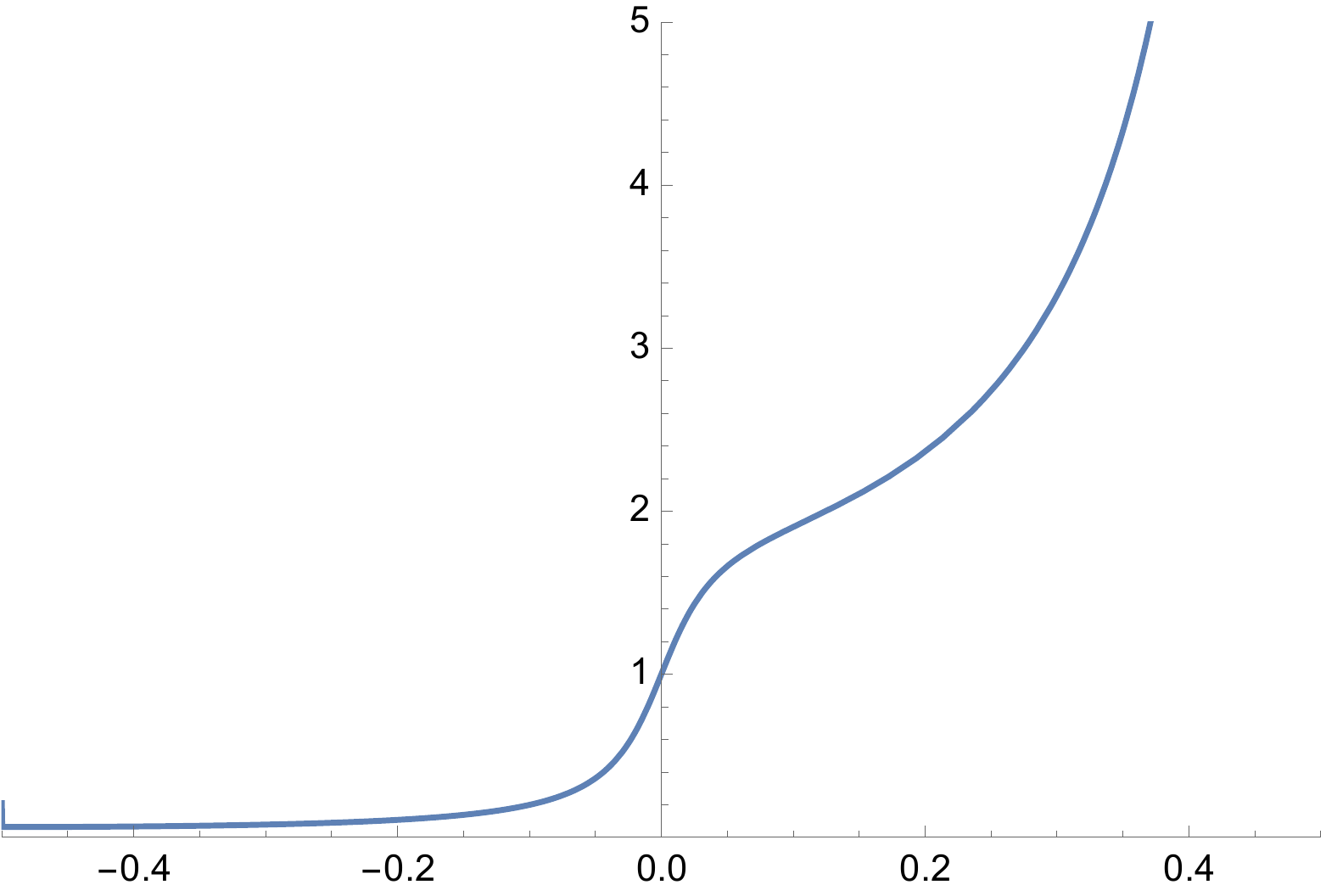}
    \caption{$(\Delta \phi/\pi)$ as a function of $(\xi/\pi)$ for $\rho = 0.1$. Here, $\xi > 0$ represents an attractive interaction, while $\xi <0$ represents a repulsive interaction.
    }
    \label{fig:Darwin-plot}
\end{figure}

It is useful to replace $E$ and $L$ by dimensionless parameters as
\begin{align}
    E = m \cosh\rho \,,
    \quad 
    \alpha = L \sin\xi \,.
\end{align}
The net change of $\phi$ between the two asymptotic $r\rightarrow \infty$ regions can be written as
\begin{align}
    \cos\left(\frac{\Delta\phi}{2} \cos\xi \right) = - \frac{\cosh\rho \sin\xi}{\sqrt{\sinh^2\rho + \sin^2 \xi}} \,.
\end{align}
Equivalently, we may write 
\begin{align}
    \frac{\Delta\phi}{2} \cos\xi = \frac{\pi}{2} + \arctan\left(\frac{\tan\xi}{\tanh\rho}\right) \,.
    \label{Delta-phi-KOT}
\end{align}
The scattering angle in our convention is $\theta = \pi -\Delta \phi$, so that $\theta > 0 $ means a repulsive interaction, 
while $\theta < 0$ means an attractive interaction. 
In the repulsive case, $\theta$ remains smaller than $\pi$. In the attractive case, the trajectory can wind around the nucleus multiple times and make $\theta$ arbitrarily negative; see Fig~\ref{fig:Darwin-plot}.

\paragraph{Radial action}  
We use the form of the radial action in \eqref{Coulomb-like-RA} with $V(r) = - \alpha/r$. We need to evaluate the integral, 
\begin{align}
        \int p_r \, dr = \int \left[ \left( E + \frac{\alpha}{r} \right)^2 - \frac{L^2}{r^2} - m ^2 \right]^{1/2} dr
        = \int \left[ 1  + \frac{2\eta }{z} - \frac{\nu^2}{z^2} \right]^{1/2} dz \,, 
\end{align}
where we introduced short-hand notations $z = kr$, $\nu = \sqrt{L^2-\alpha^2} = L \cos\xi$, $\eta = E\alpha/k$. 
Regulating and evaluating the integral, we obtain 
\begin{align}
\begin{split}
       I(k,L) &= 2 \lim_{R\rightarrow \infty} \left( \int_{r_\mathrm{min}}^R p_r dr - k R - \eta \log(2kR) \right)
       \\
       &= -\nu\pi - 2 \nu \arctan \left(\frac{\eta}{\nu}\right) -  \eta\left[\log \left(\nu^2+\eta^2\right)-2\right] \,.
       \label{Coulomb-RA-final}
\end{split}
\end{align}
As a consistency check, inserting this into \eqref{Delta-phi-RA}, we recover \eqref{Delta-phi-KOT}.

\paragraph{In and out states} 

We start with the time-independent Klein-Gordon equation, 
\begin{align}
   \left[ \left( E + \frac{\alpha}{r} \right)^2 + \nabla^2 - m^2 \right] \psi(\vec{r}) = 0 \,, 
   \quad 
   E = \sqrt{k^2 + m^2} \,.
\end{align}
We assume the partial wave ansatz \eqref{in-partial}:
\begin{align}
    \psi_\mathrm{in}(\vec{k} ;\vec{r}) =  \sum_{n=0}^\infty (2n+1) i^n  e^{i\delta_n} R_n(kr) \l( r|k \r)^n_{0,0} \,.
    \nonumber 
\end{align}
The radial equation for $L_n(z=kr) = z R_n(z)$ is 
\begin{align}
    z^2 L_n''(z) +\left[ z^2+2 \eta z - \nu(\nu+1)\right] L_n(z) =0 \,,
    \label{Coulomb-radial-eq}
\end{align}
where the parameters $\eta$ and $\nu$ are defined as 
\begin{align}
    \eta = E\alpha/k \,,
    \quad 
    \nu (\nu+1) = n(n+1) - \alpha^2 \,.
\end{align}

Up to a mapping of parameters, the relativistic Coulomb equation \eqref{Coulomb-radial-eq} take the same form as the non-relativistic one, which in turn is identified with the Whittaker equation. 
In our convention, the normalized radial wave-function is given by 
\cite{landau1981quantum,Gaspard:2018xgb}
\begin{align}
    R_n(z) = \frac{e^{\pi \eta / 2}|\Gamma(\nu-i \eta+1)|}{\Gamma(2 \nu+2) } (2z)^{\nu} e^{i z} M(\nu-i \eta+1, 2 \nu+2,-2 i z) \,,
    \label{coulomb-Rn}
\end{align}
where $M(a,b,z) = {}_1F_1(a;b;z)$ is Kummer's confluent hypergeometric function. Despite its appearance, $R_n(z)$ is real-valued for real $z$. The large $z$ limit of $R_n(z)$ is known to be
\begin{align}
    R_n(z) \approx \frac{1}{z} \sin\left[ z +\eta \log(2z) -\frac{\nu}{2}\pi + \arg\Gamma(\nu-i\eta+1) \right] \,.
    \label{coulomb-phase}
\end{align}

\paragraph{S-matrix, phase shift, scattering angle}

Comparing \eqref{coulomb-phase} to the general definition of the phase shift $\delta_n$ in \eqref{delta-n-def} 
and disregarding the famous $\log(2z)$ tail, 
we obtain 
\begin{align}
    2 \delta_n =  (n-\nu)\pi + 2 \arg\Gamma(\nu-i\eta+1) \,.
\end{align}
In the large $n$ limit, we can use Stirling's approximation for the $\Gamma$ function to find 
\begin{align}
    2\delta_n \approx (n-\nu)\pi - 2 \nu_+ \arctan \left(\frac{\eta}{\nu}\right) -  \eta\left[\log \left(\nu^2+\eta^2\right)-2\right] \,.
    \label{Coulomb-phase-final}
\end{align}
We define the classical limit by taking $n \rightarrow \infty$, $\alpha\rightarrow \infty$, $\hbar\rightarrow 0$ while keeping fixed  
\begin{align}
    L_\mathrm{cl} = n_+ \hbar \,, \quad \alpha_\mathrm{cl} = \alpha \hbar \,,
    \quad \nu_\mathrm{cl} = \nu_+ \hbar  = (\nu+1/2)\hbar \,.
\end{align}
Comparing \eqref{Coulomb-phase-final} and \eqref{Coulomb-RA-final}, 
we confirm that $(2\delta_n)\hbar$ converges to $\pi L + I(k,L)$ such that 
\begin{align}
    \theta = \left.  2 \frac{d \delta_n}{dn} \right|_\mathrm{cl} \,.
    \label{theta-cl}
\end{align}
This relation will play a central role in the coming subsections.

\subsection{KMOC via Clebsch-Gordan} \label{sec:KMOC-CG}

In this and the next subsections, 
we explain how to combine the KMOC formula with the partial wave expansion.
For simplicity, we focus on the scalar case. 

\paragraph{Classical impulse from classical mechanics} 

Classically, the conservation of orbital angular momentum dictates 
that the impulse lies in the plane spanned by $\vec{k}$ and $\vec{b}$:
\begin{align}
    \Delta \vec{k} = (\cos\theta -1) \vec{k} + \sin\theta |\vec{k}| \hat{b} \,.
\end{align}
The scattering angle $\theta$ is a function of $|\vec{k}|$ and $|\vec{b}|$. 

For a later purpose, it is useful to take the Fourier transform of the components. 
\begin{align} \label{KMOC-fourier-0}
    \Delta \vec{k} = |\vec{k}| \int \hat{d}^2 q\, e^{i\vec{q}\cdot\vec{b}} \, \left( F_0(\vec{q}) \hat{k} + F_2(\vec{q}) \hat{q} \right) \,.
\end{align}
Projecting out components, taking the inverse transform, 
and using the defining property of the Bessel function with $\hat{b}\cdot\hat{q} = \cos\phi$, 
\begin{align}
    \int d\phi \, e^{-i z \cos\phi} e^{in\phi} = 2\pi (-i)^n J_n(z) \,,
\end{align}
we find 
\begin{align} \label{KMOC-fourier}
\begin{split}
      F_0 &= \int d^2b \, e^{-i\vec{q}\cdot\vec{b}}  (c_\theta-1) = 2 \pi \int_{0}^{\infty} db \, b \, (c_\theta-1) \, J_0 \left(qb \right) \,,
    \\
    F_2 &= \int  d^2b \, e^{-i\vec{q}\cdot\vec{b}} (\hat{q}\cdot \hat{b}) s_\theta = - 2 \pi i \int_{0}^{\infty} db\, b \,s_\theta \, J_1 \left(qb \right) \,.
\end{split}
\end{align}
In the last step, to avoid clutter, we replaced $|\vec{b}|$ by $b$ and $|\vec{q}|$ by $q$. 

\paragraph{Classical impulse from quantum amplitude}

Recall the impulse formula \eqref{KMOC-full-CM}, 
\begin{align}
\begin{split}
         \Delta \vec{k}  &= \frac{\hbar^2}{|\vec{k}|^2} \int \hat{d}^2 q \,e^{i \vec{q}\cdot \vec{b} } \,\vec{K} (\vec{k},\vec{q}) \,,
         \\
         \vec{K}(\vec{k},\vec{q})  &= \int \hat{d}^2 \hat{k}_2 \left[ S^\dagger(\vec{k}_3,\vec{k}_2) (\vec{k}_2 - \vec{k}) S(\vec{k}_2, \vec{k}_1) \right] \,,
         \quad 
         \vec{k}_3  = \vec{k} + \hbar \vec{q}/2\,, \quad \vec{k}_1 = \vec{k} - \hbar \vec{q}/2\,.
\end{split}
\label{KMOC-short-CM}
\nonumber 
\end{align}
Recall also the partial wave expression \eqref{S-partial-scalar} for the S-matrix, 
\begin{align}
     S(\vec{k}', \vec{k}) = \pi \sum_{n=0}^\infty (2n+1) e^{2i\delta_n} \l( k'| k \r)^n_{0,0} \,.
\nonumber 
\end{align}

The well known recursion relation for Legendre polynomials, 
\begin{align}
  (2 n+1) x P_n(x) =   (n+1) P_{n+1}(x) + n P_{n-1}(x) \,,
\end{align}
implies a Clebsch-Gordan relation for the spinor products, 
\begin{align}
    (2 n+1) (\hat{k}'\cdot \hat{k}) \l( k'| k \r)^n_{0,0} =   (n+1) \l( k'| k \r)^{n+1}_{0,0} + n \l( k'| k \r)^{n-1}_{0,0} \,.
\end{align}
Applying this relation to $k = k_\pm$, we can compute 
\begin{align} \label{KMOC-k}
\begin{split}
     \vec{k} \cdot \vec{K} &= \frac{1}{2} (\vec{k}_3 + \vec{k}_1) \cdot \vec{K}
     \\
      &= \pi |\vec{k}_3|^2 \sum_{n=0}^{\infty} (n+1) \left[ \cos 2(\delta_{n+1} -\delta_{n}) -1 \right] \left[ P_{n+1}(\alpha) + P_{n}(\alpha)  \right] 
     \\
     &\quad +4 \pi^2 \left(|\vec{k}_3|^2 -  |\vec{k}|^2\right) \delta^2\left( \frac{\hbar \vec{q}}{|\vec{k}|} \right) \,,
\end{split}
\end{align}
where $\cos \alpha = \hat{k}_3 \cdot \hat{k}_1$. Similarly, 
\begin{align} \label{KMOC-q}
\begin{split}
     \hbar \vec{q} \cdot \vec{K} & = (\vec{k}_3 - \vec{k}_1) \cdot \vec{K}
     \\
     & = 2i \pi |\vec{k}_3|^2 \sum_{n=0}^{\infty} (n+1) \sin 2(\delta_{n+1} -\delta_{n}) \left[ P_{n+1}(\alpha) - P_{n}(\alpha)   \right] 
      \,.
\end{split}
\end{align}
We can also show that, in this simple setting, 
\begin{align}
     (\vec{k} \times \vec{q}) \cdot \vec{K} &= 0 \,.
\end{align}

We are now ready to take the classical limit in the sense that
\begin{align}
    n\rightarrow \infty, 
    \quad 
    \hbar \rightarrow 0, 
    \quad 
   n_+ \hbar \equiv (n+1/2) \hbar =  |\vec{k}||\vec{b}| = \mbox{(fixed)}\,.
\end{align}
In terms of the partial wave sum, the classical limit is simply a continuum limit,  
\begin{align}
    \sum_n \;\; \rightarrow \;\; \int dn_+ = \frac{|\vec{k}|}{\hbar} \int db  \,.
\end{align}
The factors in the ``integrand" become 
\begin{align}
    n+1 \approx  \, \frac{|\vec{k}|}{\hbar} |\vec{b}| \,, 
    \qquad
    \delta_{n+1} -\delta_{n}  \approx \left. \frac{d \delta_n}{dn} \right|_\mathrm{cl} \equiv \frac{1}{2} \theta_\mathrm{cl} \,.
\end{align}
The most non-trivial step of the classical limit concerns $P_n(\alpha)$. 
As we observed in \eqref{small-alpha}, it is sufficient to focus on the small angle limit $(\alpha \ll 1)$, where 
\begin{align}
    P_n(\alpha) \approx J_0(n_+ \alpha) \approx J_0 (qb ) \,.
\end{align}
Then the sum in \eqref{KMOC-k} becomes
\begin{align} \label{K-sum-k}
\begin{split}
    & \sum_{n=0}^{\infty} (n+1) \left[ \cos 2(\delta_{n+1} -\delta_{n}) -1 \right] \left[ P_{n+1}(\alpha) + P_{n}(\alpha)  \right] 
    \\
    & \rightarrow \quad \frac{2|\vec{k}|^2}{\hbar^2} \int_{0}^{\infty} db\, b 
    \left( \cos\theta_\mathrm{cl} - 1 \right) J_0 \left(qb \right) \,. 
\end{split}
\end{align}
The delta-function term in \eqref{KMOC-k} is $\mathcal{O}(\hbar^0)$ and does not contribute to the classical impulse. 
The sum in \eqref{KMOC-q} is slightly more interesting. Note that 
\begin{align}
    P_{n+1}(\alpha) - P_{n}(\alpha) \approx \frac{d}{dn_+} J_0(n_+ \alpha) = -\alpha \, J_1(n_+ \alpha) = - \frac{\hbar|\vec{q}|}{|\vec{k}|} J_1 \left(qb \right) \,.
\end{align}
Then, the sum becomes
\begin{align} \label{K-sum-q}
\begin{split}
    & \sum_{n=0}^{\infty} (n+1) \sin 2(\delta_{n+1} -\delta_{n}) \left[ P_{n+1}(\alpha) - P_{n}(\alpha)  \right] \\
    & \rightarrow \quad -\frac{\hbar|\vec{q}|}{|\vec{k}|} \frac{|\vec{k}|^2}{\hbar^2} \int_{0}^{\infty} db \, b\,   
    (\sin \theta_\mathrm{cl} ) \, J_1 \left(qb\right) \,.
\end{split}
\end{align}

Let us summarize what we have done so far. From \eqref{KMOC-k} and \eqref{K-sum-k}, we find
\begin{align}
    (\hat{k} \cdot \vec{K}) |_\mathrm{cl} = \frac{2\pi |\vec{k}|^3 }{\hbar^2} \int_{0}^{\infty} db\, b 
    \left( \cos\theta_\mathrm{cl} - 1 \right) J_0 \left(qb \right) \,.
\end{align}
From \eqref{KMOC-q} and \eqref{K-sum-q}, we find 
\begin{align}
    (\hat{q} \cdot \vec{K}) |_\mathrm{cl} = -\frac{2\pi i |\vec{k}|^3 }{\hbar^2}  \int_{0}^{\infty} db \, b\,   
    (\sin \theta_\mathrm{cl} ) \, J_1 \left(qb\right) \,.
\end{align}
These formulas agree perfectly with their classical counterparts \eqref{KMOC-fourier-0}, \eqref{KMOC-fourier}:
\begin{align}
      \frac{1}{k} (\hat{k}\cdot\vec{K})|_\mathrm{cl} = \frac{k^2}{\hbar^2} F_0  \,,
        \quad 
     \frac{1}{k}    (\hat{q}\cdot\vec{K})|_\mathrm{cl} = \frac{k^2}{\hbar^2} F_2 \,. 
\end{align} 
%

\subsection{KMOC via saddle point} \label{sec:KMOC-saddle}

\paragraph{Saddle point method review} 

For simplicity, we focus on the scalar scattering. 
The way the saddle point method works is similar 
to how to evaluate a partition function in statistical mechanics. 
{}We regard the S-matrix as a (complex-valued) partition function, 
\begin{align}
    S(k,\theta)  = 2\pi \sum_{n=0}^\infty n_+ e^{2i\delta_n} P_n(\cos\theta) \,.
\end{align}
We also make some simplifying assumptions:
\begin{align}
    \delta_{n+1} - \delta_n > 0 \,,
    \quad 
    \delta_{n+1} - 2\delta_n + \delta_{n-1} <0 \,.
\end{align}
Physically, it means that the scattering is repulsive, so that only the  winding number sector contributes. 
We expect that a generalization to a non-zero winding number will be  straightforward. 

For a fixed value of $\theta$, the dominant contribution to the sum 
can be identified from the asymptotic formula for $n_+ \gg 1$, 
\begin{align}
\begin{split}
       e^{2i\delta_n} P_n(\cos\theta) &\approx e^{2i\delta_n} \left(\frac{\theta}{\sin\theta}\right)^{1/2}J_0(n_+\theta) 
       \\
       &\approx \frac{1}{\sqrt{2\pi n_+\sin\theta}} \left( e^{2i\delta_n -i n_+\theta + \pi i/4 } +\cdots \right) \,.
\end{split}
\end{align}
It follows that the saddle point condition is 
\begin{align}
    \theta = \frac{d}{dn} (2\delta_n) \approx 2(\delta_{n+1} - \delta_n) \,.
    \label{saddle-theta-recap}
\end{align}

Pushing the statistical mechanics analogy further, 
we can compare the two ways to compute the expectation value of $n_+$: 
\begin{align}
\begin{split}
     \langle n_+ \rangle &= \frac{2\pi}{S} \sum_n n_+^2 e^{2i\delta_n} P_n(\cos\theta) \,,
\\
    [ n_+ ] &= \frac{2\pi}{S} \sum_n n_+ e^{2i\delta_n} \left[ i \partial_\theta P_n(\cos\theta) \right] = S^{-1} (i\partial_\theta S)  \,.
\end{split}
\label{n-expectation}
\end{align}
The two values are not strictly equal, 
but in the classical limit, they converge to the same value\footnote{
It is amusing to test this statement numerically. 
The partial wave sum converges extremely slowly, 
so an algorithm to accelerate convergence, such as the Shanks transformation \cite{https://doi.org/10.1002/sapm19553411,bender1999advanced}, is strongly recommended. 
} 
with a relative error scaling as some negative power of $\langle n_+\rangle$.
This expectation value leads to a way to reproduce the classical relation $\theta(b)$, alternative to \eqref{saddle-theta-recap}, 
\begin{align}
   kb_*(\theta) = S^{-1} (i\hbar \partial_\theta S)  \,.
   \label{saddle-dual} 
\end{align}

The final result for the saddle point approximation is 
\begin{align}
    S(\theta) \approx \frac{2\pi k}{\hbar} \left( \frac{b_*(\theta) |b_*'(\theta)|}{\sin\theta} \right)^{1/2} \exp\left[ i Q(\theta)/\hbar + i\lambda \right] \,.
    \label{BM-saddle} 
\end{align} 
Here, $Q(k,\theta)$ is a Legendre transform of the radial action 
as in \eqref{I-Q-Legendre}: 
\begin{align}
    Q(k,\theta) = \widehat{I}(k,k b_*(\theta)) - k b_*(\theta) \theta \,, 
\end{align}
and $b_*(\theta)$ is determined by \eqref{theta-IL} as 
\begin{align}
    \theta = \left. \frac{\partial \widehat{I}}{\partial L} \right|_{L=kb_*(\theta)} \,.
\end{align}
The prefactor of \eqref{BM-saddle} and the phase $\lambda$ are all fixed by the saddle point evaluation of the integral which originates from the continuum limit of the sum over $n$. 
The phase $\lambda$ is independent of $\theta$, so it will not affect physical observables.

\paragraph{KMOC from the saddle point}

We are ready to apply the saddle point approximation to the KMOC formula.  
\begin{align}
\begin{split}
         \vec{k}_\mathrm{out}  &= \frac{\hbar^2}{k^2} \int \hat{d}^2 q \,e^{i \vec{q}\cdot \vec{b} } \,\vec{K}_\mathrm{out} (\vec{k},\vec{q}) \,,
         \\
         \vec{K}_\mathrm{out}(\vec{k},\vec{q})  &= \int \hat{d}^2 k_2 \left[ S^\dagger(\vec{k}_3,\vec{k}_2) \vec{k}_2 S(\vec{k}_2, \vec{k}_1) \right] \,,
         \quad
         \vec{k}_3 = \vec{k} + \hbar \vec{q}/2\,, \quad \vec{k}_1 = \vec{k} - \hbar \vec{q}/2\,.
\end{split}
\nonumber
\label{KMOC-full-CM-copy-2} 
\end{align}

The displacement $\vec{k} \rightarrow \vec{k} - \hbar \vec{q}/2$ changes the angle $\theta$ by 
\begin{align}
    \cos(\theta + \delta\theta) \approx \frac{\vec{k}_2\cdot (\vec{k} - \hbar \vec{q}/2)}{|\vec{k}|^2} 
    \quad \Longrightarrow \quad 
    \sin\theta \,\delta \theta = \frac{\hbar}{2k} \hat{k}_2\cdot \vec{q}  \,. 
\end{align}
We use the relation \eqref{saddle-dual} to get 
\begin{align}
    S(\vec{k}_2, \vec{k}_1) \approx S(\vec{k}_2, \vec{k}) \exp\left[ \frac{-ib_*(\theta)}{2\sin\theta } \hat{k}_2\cdot \vec{q} \right] \,.
\end{align}
Similarly, we find
\begin{align}
    S^\dagger( \vec{k}_3,\vec{k}_2) \approx S^\dagger( \vec{k},\vec{k}_2) \exp\left[ \frac{-ib_*(\theta)}{2\sin\theta } \hat{k}_2\cdot \vec{q} \right] \,.
\end{align}
Combining the two, we obtain 
\begin{align}
         \vec{K}_\mathrm{out}(\vec{k},\vec{q})  &\approx \int \hat{d}^2 k_2 \, \vec{k}_2 \, |S(\vec{k}_2, \vec{k}_1) |^2  \exp\left[ \frac{-ib_*(\theta)}{\sin\theta } \hat{k}_2\cdot \vec{q} \right]  \,.
\label{KMOC-saddle} 
\end{align}
Next, we perform the $q$-integral, which produces the delta-function, 
\begin{align}
    \delta^2\left( \frac{\hat{k}_{2\perp}}{\sin\theta} b_*(\theta) - \vec{b} \right) \,.
    \label{q-to-delta}
\end{align}
In general, we can write 
\begin{align}
    \hat{k}_2 = \hat{k} \cos\theta + \hat{k}_{2\perp} = \hat{k} \cos\theta  + \hat{b} \sin\theta \cos\phi  + (\hat{k}\times \hat{b}) \sin\theta \sin\phi \,. 
\end{align}
Now the $\hat{k}_2$ integral is completely localized by the delta-function, 
\begin{align}
  \int \frac{\sin\theta d\theta d\phi}{(2\pi)^2} \,   \delta^2\left( \frac{\hat{k}_{2\perp}}{\sin\theta} b_*(\theta) - \vec{b} \right)
  = \frac{1}{(2\pi)^2} \left. \frac{\sin\theta}{b_*(\theta) |b_*'(\theta)|} \right|_{b*(\theta) = b} \,.
  \label{k-prime-scalar}
\end{align}

Finally, combining \eqref{BM-saddle} and the $|S|^2$ factor in \eqref{KMOC-saddle}, we confirm that 
\begin{align}
    \hat{k}_2(k,b)  =\left[ \cos \theta_*(k,b) \right] \hat{k} + \left[ \sin \theta_*(k,b) \right] \hat{b} \,.
\end{align}
In summary, the KMOC formula is guaranteed to reproduce the classical impulse, provided that the partial wave phase converges to the radial action in the classical limit, and that the saddle point is isolated and non-degenerate.

\section{Charge-monopole scattering} \label{sec:charge-monopole} 

\begin{figure}[ht]
    \centering
    \includegraphics[width=10cm]{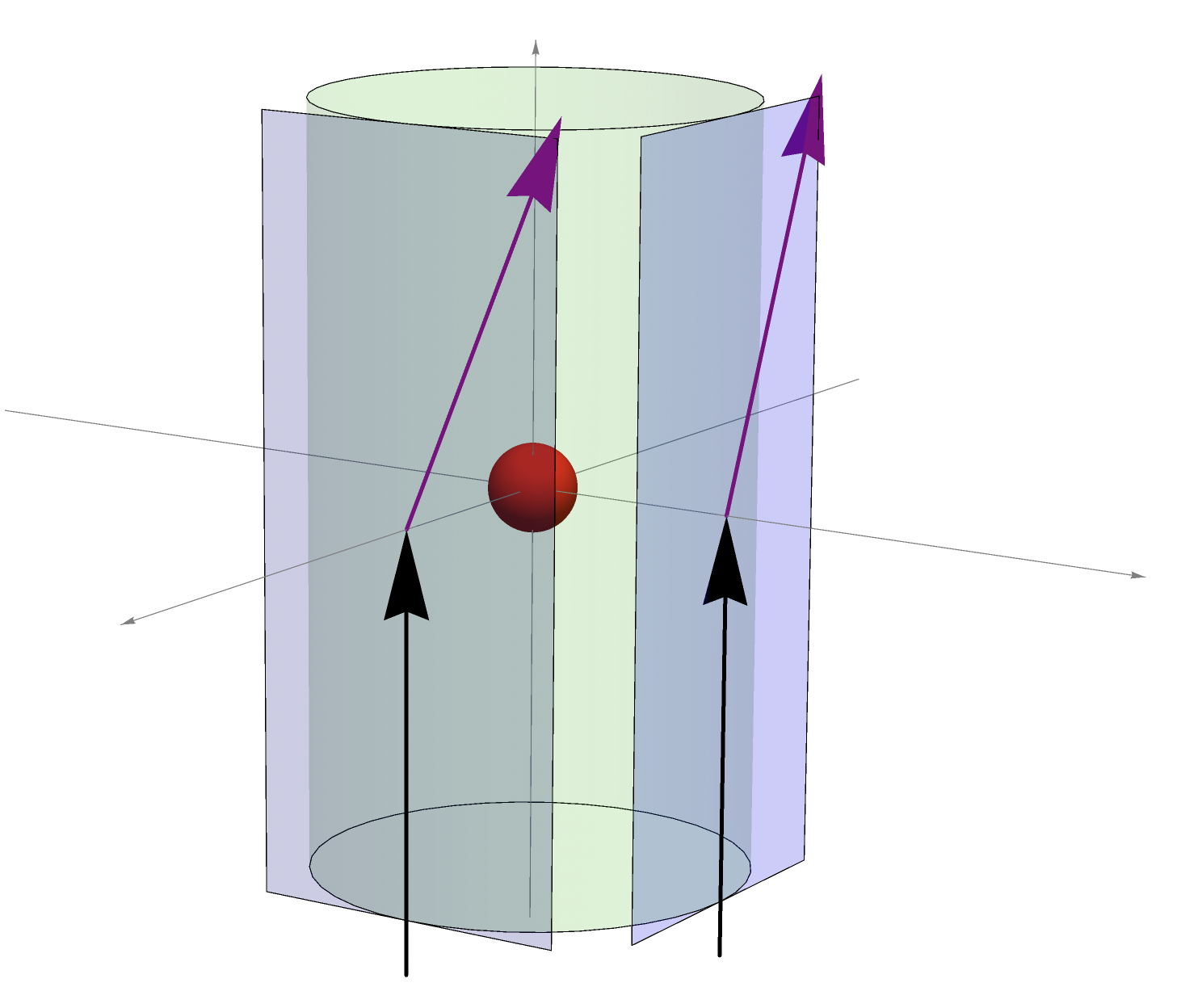}
    \caption{A schematic description of the charge-monopole scattering in the probe limit.
    The initial momentum is in the $\hat{z}$-direction, the leading order impulse is in the $\hat{\phi}$-direction.
    }
    \label{fig:charge-monopole-schematics}
\end{figure}

As a simple but non-trivial example of a not purely scalar scattering problem, 
we revisit the charge-monopole scattering. 
In the probe limit, both the classical equation of motion and the quantum wave equation are exactly solvable \cite{Schwinger:1976fr,Boulware:1976tv}. 
Compared to the Coulomb scattering, a notable difference is that the classical impulse is characterized by 
two scattering angles (as opposed to one), 
as indicated in in Fig.~\ref{fig:charge-monopole-schematics}.
To the best of our knowledge, how to reproduce both angles from a quantum amplitude has not been discussed in the literature. 
We fill this gap by combining our spinor-based partial wave expansion and the KMOC formula. 

\subsection{Classical impulse} 

\paragraph{Trajectory}

We begin with a brief review of the classical trajectory. 
As explained in \cite{Schwinger:1976fr,Boulware:1976tv}, 
the non-relativistic problem is exactly solvable. 
In the probe limit, the relativistic problem differs from the non-relativistic one only through 
the definition of the momentum. So, the classical trajectory remains essentially the same. 
The conserved energy is simply 
\begin{align}
    E = \sqrt{k^2 + m^2} \,, 
    \quad k = |\vec{k}| \,.
\end{align}
In this section, we denote the asymptotic incoming momentum by $\vec{k}$ 
and the dynamical (non-constant) momentum by $\vec{p}$. 
The conserved angular momentum 
includes the orbital and field contributions: 
\begin{align}
        \vec{\mathcal{J}} = \vec{L} + \vec{M} \,, 
        \quad 
        \vec{L} = \vec{r} \times \vec{p} \,,
        \quad 
        \vec{M} = - \frac{eg}{4\pi} \hat{r} = - h_{\mathrm{cl}} \hat{r} \,. 
\end{align}
A peculiar feature of this problem is that the trajectory of the probe particle is confined on a cone: 
\begin{align}
    \hat{\mathcal{J}} \cdot \hat{r} = - \sin\xi \,,
    \quad 
    \tan\xi 
    =  \frac{h_{\mathrm{cl}}}{|\vec{k}| |\vec{b}|} \equiv \kappa \,.
    \label{xi-def}
\end{align}
The relativistic kinematics enters only through the definition of the 4-momentum. Otherwise, the relativistic problem is the same non-relativistic one. In particular, the scattering angles depend only on $\xi$ defined in \eqref{xi-def} but not on $k$.

\paragraph{Radial action}

Refs.~\cite{Schwinger:1976fr,Boulware:1976tv} showed how to map the classical trajectory on the cone to the plane orthogonal to $\vec{\mathcal{J}}$. Let $(\rho, \psi)$ be the polar coordinates of the plane. The dynamics is encoded by an effective Hamiltonian which leads to a relation,
\begin{align}
    p_\rho^2 + \frac{J^2-h^2}{\rho^2} = k^2 \,.
\end{align}
The radial action for $\rho$ gives the deflection of $\psi$. 
\begin{align} 
\begin{split}
     I(k,J) &= 2 \lim_{\rho_\mathrm{max} \rightarrow \infty} \left( \int_{\rho_\mathrm{min}}^{\rho_\mathrm{max}} p_\rho \,d\rho - k \rho_\mathrm{max} \right)
    = -\pi \sqrt{J^2-h^2} \,. 
\\
    &\Longrightarrow\quad \Delta \psi = - \frac{\partial I}{\partial J} = \pi \frac{J}{\sqrt{J^2-h^2}} = \frac{\pi}{\cos\xi} \,. 
\end{split}
\label{Delta-psi}
\end{align}

\paragraph{Impulse}

Let $\vec{k}$ and $\vec{k}'$ be the initial and final asymptotic momenta, 
and $\Delta \vec{k} = \vec{k}'-\vec{k}$. 
The exact impulse formula can be written as 
\begin{align}
    \Delta \vec{k} =  (c_\theta - 1) \vec{k} + c_2 |\vec{k}| \hat{b}
    +  c_1 (\vec{k} \times \hat{b}) \,.
    \label{impulse-non-pert}
\end{align}
Using the cone-to-plane map of \cite{Schwinger:1976fr,Boulware:1976tv}, one can show that the scattering angle $\theta$ defined by $\hat{k}\cdot\hat{k}' = \cos\theta$  is related to $\xi$ and $\Delta \psi = \pi/\cos\xi$ non-linearly as 
\begin{align}
\begin{split}
    &\cos\theta = - \cos^2\xi \cos \left(\frac{\pi}{\cos\xi}\right) - \sin^2\xi \,.
    \\
\Longrightarrow \quad &\cos^2(\theta/2) = \cos^2\xi \sin^2\left(\frac{\pi}{2\cos\xi}\right) \,.
\end{split}
    \label{scattering-angle-BBCEL-CDMST}
\end{align}
Th other two coefficients can be determined by 
\begin{align}
\begin{split}
      & (\vec{k} + \vec{k}')\cdot \vec{\mathcal{J}} 
      = 0 \,,
    \quad  c_1^2+c_2^2 = \sin^2\theta 
    \\
    \Longrightarrow
    \quad 
    & c_1 = \sin \xi \cos\xi \left[1 -  \cos\left( \frac{\pi}{\cos\xi} \right) \right]\,,
    \quad 
    c_2 = \cos\xi \sin\left( \frac{\pi}{\cos\xi} \right) \,.
\end{split}
\label{c1_c2}
\end{align}
The sign of $c_2$ is fixed by the fact that $\Delta \psi = \pi/\cos\xi > \pi$. 

Perturbatively, we can compute $\Delta \vec{k}$ order by order in $\kappa \approx \xi \approx \theta/2$. To the leading order in $\kappa$, we obtain 
\begin{align}
    \Delta \vec{k}^{(1)} = 2\kappa (\vec{k} \times \hat{b}) \,.
    \label{impulse-bbcel-leading}
\end{align}
At the $\kappa^2$ order, we find 
\begin{align}
\Delta \vec{k}^{(2)} = -2\kappa^2 \vec{k} -\frac{\pi}{2} \kappa^2 |\vec{k}| \hat{b}  \,.
\label{impulse-bbcel-sub-leading}
\end{align}
One can try to reproduce these results starting form a classical perturbation theory; see appendix~\ref{app:pert-charge-momopole}. Amusingly, the perturbative computation turns out to be substantially more complicated than solving the equation of motion exactly

\paragraph{Fourier transform} 
As before, we consider the Fourier transform of the non-perturbative impulse formula \eqref{impulse-non-pert}:
\begin{align}
\begin{split}
    \Delta \vec{k} = |\vec{k}| \int \hat{d}^2 q \,  e^{i\vec{b}\cdot\vec{q}} \, 
    \left( F_0 \hat{k} + F_2 \hat{q} + F_1 (\hat{k} \times \hat{q}) \right) \,,
    \quad
    F_i = F_i(\vec{k}, \vec{q}, h_{\mathrm{cl}})\,,
    \quad  h_{\mathrm{cl}} = \frac{eg}{4\pi} \,.
\end{split}
\end{align}
Taking the inverse Fourier-transform and projecting out components, we find 
\begin{align}
\begin{split}
     F_0 = \int d^2 b \, e^{-i\vec{b}\cdot\vec{q}} (c_\theta-1) &= \frac{2 \pi h_{\mathrm{cl}}^2}{|\vec{k}|^2} \int_0^{\infty} dx \, x (c_\theta-1) J_0 \left(\frac{h_{\mathrm{cl}} |\vec{q}|}{|\vec{k}|} x \right)  \,, 
     \\
     F_2 = \int d^2 b \, e^{-i\vec{b}\cdot\vec{q}}  (\hat{b} \cdot \hat{q} ) \, c_2   &=  - i \frac{2 \pi h_{\mathrm{cl}}^2}{|\vec{k}|^2} \int_0^{\infty} dx \, x \, c_2 \, J_1 \left(\frac{h_{\mathrm{cl}} |\vec{q}|}{|\vec{k}|} x \right) \,,
     \\
     F_1 =  \int d^2 b \, e^{-i\vec{b}\cdot\vec{q}}  (\hat{b} \cdot \hat{q})\, c_1  &=  - i \frac{2 \pi h_{\mathrm{cl}}^2}{|\vec{k}|^2}\int_0^{\infty} dx \, x \, c_1 \, J_1 \left(\frac{h_{\mathrm{cl}} |\vec{q}|}{|\vec{k}|} x \right)\,.
\end{split}
\label{G-classical-bessel}
\end{align}
where $x= \cot\xi$ comes from the definition of $\xi$ in \eqref{xi-def}.

\subsection{Wave equation and S-matrix} 

\paragraph{Wave equation} 
In the probe limit, the only difference between the Schrödinger equation and the Klein-Gordon equation 
is the $E(\vec{k})$ dispersion relation. In both cases, the wave equation reads 
\begin{align}
\vec{D}^2 \psi(\vec{r}) = 
    \left( \nabla - i e \vec{A}\right)^2 \psi(\vec{r}) = - k^2 \psi(\vec{r}) \,.
    \label{plane-wave-eq-copy}
\end{align}
The magnetic field of the monopole is assumed to be 
\begin{align}
    \vec{B} = \frac{g}{4\pi} \frac{\hat{r}}{r^2} \,.
\end{align}
The vector potential is given by 
\begin{align}
\begin{split}
    \vec{A}_N &= \frac{g}{4\pi} \frac{\hat{z}\times \hat{r}}{r+z} = \frac{g}{4\pi r} \left(\frac{ 1-\cos\theta}{\sin\theta}\right) \hat{\phi}\,,
    \\
    \vec{A}_S &= - \frac{g}{4\pi} \frac{\hat{z}\times \hat{r}}{r-z} = \frac{g}{4\pi r} \left(\frac{-1-\cos\theta}{\sin\theta}\right) \hat{\phi}  \,.
\end{split} 
\label{vect-pot-NS}
\end{align}
The gauge transformation between the two hemispheres is given by
\begin{align}
\begin{split}
    e \vec{A}_N - e \vec{A}_S &= \frac{eg}{2\pi} \frac{\hat{\phi}}{r\sin\theta} = \nabla (2h \phi) \,,
    \quad 
    h = \frac{eg}{4\pi} \,,
    \quad
    \psi_N /\psi_S = e^{+2ih \phi} \,.
\end{split}
\label{monopole-gauge-transf}
\end{align}
The total angular momentum operator satisfies
\begin{align}
    \vec{\mathcal{J}} = \vec{r} \times \left( -i\vec{D} \right) - h\,\hat{r} 
    \quad 
    \Longrightarrow
    \quad 
    \vec{\mathcal{J}}^2 - h^2 = r^2 (-i\vec{D})^2 + (\vec{r}\cdot\nabla)^2 + \vec{r}\cdot\nabla\,.
    \label{jj-monopole}
\end{align}
It follows that the wave equation can be solved by separation of variables.
\begin{align}
\begin{split}
        &\psi(\vec{r}) = R(kr) Y(\theta,\phi) \,,
        \quad
        \vec{\mathcal{J}}^2 Y = n(n+1)Y \,,
        \\
        &\left( - \frac{1}{r} \partial_r^2 r +\frac{n(n+1)-h^2}{r^2}\right) R  = k^2 R \,.
        \label{sep-var}
\end{split}
\end{align}
The angular part $Y$ is spanned by the monopole spherical harmonics $Y_{h,m}^n$ or equivalently Wigner D-matrices. We prefer using the spinor products introduced in section~\ref{sec:helicity}.
The normalized radial wave function is the spherical Bessel function $j_{l}(kr)$ with the index $l$ being the positive root of 
\begin{align}
    l(l+1) = n(n+1) - h^2 \,.
    \label{l-vs-n}
\end{align}

\paragraph{Asymptotic in and out states}

The in-states can be written as
\begin{align}
        \psi_\mathrm{in}(\vec{k} ;\vec{r})_h = e^{i \pi h } \sum_{n=|h|}^\infty (2n+1) i^n e^{i\pi (n-l)/2}  j_{l}(kr) \l( r|k \r)^n_{-h,h} \,.
\label{plane-wave-in}
\end{align}
The phase shift, $\delta_n = \pi (n-l)/2$, follows from the asymptotic formula for $j_l(kr)$ in \eqref{sph-bessel-asymp} generalized to non-integer indices. The gauge transformation \eqref{monopole-gauge-transf} requires that the angular wave-function should be $\l( r|k \r)^n_{-h,h}$.
The overall phase $e^{i \pi h}$ has been fixed such that the incoming wave $e^{-ikr}$ terms all have the same phase as in the partial wave expansion of the free theory. 
Similarly, the out-state can be written as 
\begin{align}
\begin{split}
      \psi_\mathrm{out}(\vec{k} ;\vec{r})_h 
        &=   \sum_{n=|h|}^\infty (2n+1) i^n e^{-i\pi (n-l)/2}  j_{l}(kr) \l( r| k \r)^n_{-h,-h} \,.
\end{split}
\label{plane-wave-out}
\end{align}
Unlike \eqref{plane-wave-in}, it does not carry an $h$-dependent overall phase.

The in-states \eqref{plane-wave-in} and the out-states \eqref{plane-wave-out} satisfy a relation of the form 
\begin{align}
    \psi_\mathrm{out}(\vec{k},\vec{r})_h = e^{ih \chi} \left[ \psi_\mathrm{in}(-\vec{k},\vec{r})_{-h} \right]^* \,.
\end{align}
The phase $\chi$ is gauge-dependent, but it will not affect the S-matrix.

By construction, the in and out states are properly normalized.  
The orthogonality \eqref{ortho-spinor} and the completeness \eqref{completeness-spinor} relations for spinors, as well as the orthogonality relation \eqref{ortho-sb} for spherical Bessel functions, leads to a simple proof of the orthogonality of the in-states.
Essentially the same proof holds for the completeness relation.

\paragraph{S-matrix}

Plugging \eqref{plane-wave-in} and \eqref{plane-wave-out} into \eqref{S-as-overlap}, and running it through \eqref{SS-vs-S}, we obtain
\begin{align}
      S(\vec{k}_2, \vec{k}_1) = \pi e^{i\pi h} \sum_{n=|h|}^{\infty} (2n+1) e^{ 2i \delta_n }  \l(k_2 |k_1 \r)^{n}_{-h,h} \,, 
      \quad 
      2\delta_n = \pi (n-l) \,.
      \label{S-matrix-exact}
\end{align}
Compared to the scalar case, it shows two notable features. 
The sum over $n$ begins at $n=|h|$ rather than 0, and 
the base amplitude is $\l(k_2 |k_1 \r)^{n}_{-h,h}$ instead of $\l(k_2 |k_1 \r)^{n}_{0,0} = P_n(\hat{k}_2\cdot \hat{k}_1)$.
The overall phase $e^{i\pi h }$ will drop out of the computation of the classical impulse.

\subsection{Clebsch-Gordan approach} 

In this subsection, we generalize the Clebsch-Gordan approach 
explained in section~\ref{sec:KMOC-CG} to the charge-monopole scattering. 
We start again with the master formula \eqref{KMOC-full-CM}:
\begin{align}
\begin{split}
         \Delta \vec{k}  &= \frac{\hbar^2}{|\vec{k}|^2} \int \hat{d}^2 q \,e^{i \vec{q}\cdot \vec{b} } \,\vec{K} (\vec{k},\vec{q}) \,,
         \\
         \vec{K}(\vec{k},\vec{q})  &= \int \hat{d}^2\hat{k}_2 \left[ S^\dagger(\vec{k}_3,\vec{k}_2) (\vec{k}_2 - \vec{k}) S(\vec{k}_2, \vec{k}_1) \right] \,,
         \\
         \vec{k}_3 & = \vec{k} + \hbar \vec{q}/2\,, \quad \vec{k}_1 = \vec{k} - \hbar \vec{q}/2 \,,
\end{split}
\nonumber
\end{align}
and the S-matrix for the charge-monopole scattering \eqref{S-matrix-exact}:
\begin{align}
      S(\vec{k}_2, \vec{k}_1) = \pi e^{i\pi h} \sum_{n=|h|}^{\infty} (2n+1) e^{ 2i \delta_n }  \l(k_2 |k_1 \r)^{n}_{-h,h} \,, 
      \quad 
      2\delta_n = \pi (n - l) \,.
      \nonumber 
\end{align}

Compared to the scalar case, a crucial difference is that $(\vec{k}\times \vec{q})\cdot \vec{K}$ no longer vanishes. 
In what follows, we will compute three terms:
\begin{align}
\begin{split}
    A_0 &= \vec{k}\cdot\vec{K} = \frac{1}{2} (\vec{k}_3 + \vec{k}_1) \cdot \vec{K} \,,
      \\ 
    A_2 &= \hbar \vec{q}\cdot\vec{K} =  (\vec{k}_3 - \vec{k}_1) \cdot \vec{K} \,,
     \\ 
    A_1 &= (\vec{k}\times \hbar \vec{q})\cdot \vec{K} =  (\vec{k}_1 \times \vec{k}_3)\cdot \vec{K} \,.
\end{split}
\label{monopole-SS}
\end{align}
Under the ``parity" operation which flips the sign of $h_{\mathrm{cl}} = e g/4\pi$, 
$A_0$ and $A_2$ are even whereas $A_1$ is odd. 
We work out the even terms first and proceed to the odd term.

\paragraph{Parity even terms}

The computation of \eqref{monopole-SS} requires Clebsch-Gordan relations in the monopole background. 
As before, our strategy is to choose a suitable relation for the Wigner d-matrices and to promote it 
to a relation for spinor products by ``covariantization".
For the parity even terms, 
the desired CG relation for d-matrices is given in \eqref{wigner-CG}. The covariant relation is then 
\begin{align}
\begin{split}
    (2n+1)(\hat{k}' \cdot \hat{k}) \l(k' |k \r)^{n}_{-h,h}
     &= f_{n+1}^h \l(k' |k \r)^{n+1}_{-h,h}  
    +f_{n}^h\l(k' |k \r)^{n-1}_{-h,h}
    - g_{n}^h \l(k' |k \r)^{n}_{-h,h}
    \,,
    \\
    f_n^h = n - \frac{h^2}{n} \,,
    &\quad 
    g_n^h = \frac{h^2}{n} + \frac{h^2}{n+1} \,.
\end{split}
\label{impulse-21-CG}
\end{align}
Using the orthogonality relation, we obtain 
\begin{align}
\begin{split}
    A_- &= \vec{k}_1 \cdot \vec{K}   
    \\
    &= \pi |\vec{k}_1|^2 \sum_{n_1,n_3} e^{2i(\delta_{n_1} -\delta_{n_3})}  \l(k_3|k_1 \r)^{n_3}_{h,h}
    \left[f_{n_1+1}^h \delta_{n_3,n_1+1} - g_{n_1}^h \delta_{n_3,n_1} + f_{n_1}^h\delta_{n_3,n_1-1} \right] 
    \\
    & \quad - |\vec{k}|^2 \hat{\delta}^2(\hat{k}_3-\hat{k}_1) \,.
\end{split}
\label{impulse-21-sum}
\end{align}
Similarly, 
\begin{align}
\begin{split}
A_+ &= \vec{k}_3 \cdot \vec{K}  
    \\
    &= \pi |\vec{k}_3|^2 \sum_{n_1,n_3} e^{2i(\delta_{n_1} -\delta_{n_3})} \l(k_3|k_1 \r)^{n_3}_{h,h}
    \left[f_{n_3+1}^h  \delta_{n_1,n_3+1} - g_{n_3}^h \delta_{n_1,n_3} + f_{n_3}^h \delta_{n_1,n_3-1} \right]
    \\
    & \quad - |\vec{k}|^2 \hat{\delta}^2(\hat{k}_3-\hat{k}_1) \,.
\end{split}
\label{impulse-32-sum}
\end{align}

Taking the ``symmetric" and ``anti-symmetric" parts, we obtain 
\begin{align}
\begin{split}
    A_0 &= \frac{1}{2} (A_+ + A_-)  = A_0^{\mathrm{(a)}} + A_0^{\mathrm{(b)}} + A_0^{\mathrm{(c)}}
    \\
    &A_0^{\mathrm{(a)}}  = - \pi |\vec{k}_3|^2 \sum_{n=|h|}^{\infty} g_n^h \l(k_3|k_1 \r)^{n}_{h,h} 
    \,,
\\
  &A_0^{\mathrm{(b)}} = - \pi|\vec{k}_3|^2 \sum_{n=|h|}^{\infty} 
    f_{n+1}^h \left[ \cos (\pi \Delta l) -1 \right] \left[\l(k_3|k_1 \r)^{n+1}_{h,h} + \l(k_3|k_1 \r)^{n}_{h,h} \right] 
    \,,
    \\
    &A_0^{\mathrm{(c)}} = - |\vec{k}|^2 \hat{\delta}^2(\hat{k}_3-\hat{k}_1) \,,
\\
    A_2 &= A_+ - A_-
    \\
    & = 2 \pi i |\vec{k}_3|^2 \sum_{n=|h|}^{\infty} f_{n+1}^h \sin (\pi \Delta l) \left[\l(k_3|k_1 \r)^{n+1}_{h,h} - \l(k_3|k_1 \r)^{n}_{h,h} \right] \,.
\end{split}
\label{three-sum}
\end{align}
We defined $\Delta l$ by 
\begin{align}
\Delta l = l' - l \,,
\quad 
(l' + 1/2)^2=(n +3/2)^2-h^2 \,.
\end{align}

\paragraph{Classical limit} 
Now, we temporarily gauge-fix the spinor products to express them as Wigner d-matrices with the relative angle between $\vec{k}_1$ and $\vec{k}_3$: 
\begin{align}
    \l(k_3|k_1 \r)^{n}_{h,h} \; \rightarrow \;  d^n_{h,h}(\alpha) \,, 
    \quad 
    \cos\alpha = \hat{k}_1 \cdot \hat{k}_3 \,.
\end{align}
As we observed earlier in the scalar case, 
the angle $\alpha$ is infinitesimal in the classical limit. In the prefactor, $|\vec{k}_+|^2$ can be safely replaced by $|\vec{k}|^2$.

The classical limit is essentially the continuum limit:
\begin{align}
    \sum_n \;\; \rightarrow \;\; \int dn_+ = h\int d\left( \frac{1}{\sin\xi} \right) \,.
\end{align}
The factors in the ``integrand" become 
\begin{align}
\begin{split}
    g_n^h = \frac{h^2}{n} + \frac{h^2}{n+1} \approx 2h \sin\xi \,,
    \quad 
    f_n^h = n - \frac{h^2}{n} \approx %
    \frac{h\cos^2\xi}{\sin\xi} \,, 
    \quad
    \Delta l \approx \frac{dl}{dn} \approx \frac{1}{\cos\xi} \,.
\end{split}
\end{align}
The most non-trivial step of the classical limit concerns $d^n_{h,h}(\alpha)$. It is sufficient to focus on the small angle limit $(\alpha \ll 1)$, where 
\begin{align}
    d^n_{h,h}(\alpha) \approx J_0(l_+ \alpha) = J_0( h\alpha \cot\xi) \,, \quad
    l_+ \equiv l + 1/2 = \sqrt{(n+1/2)^2 - h^2} \,.
\end{align}
A derivation of this approximation formula is given in appendix~\ref{app:wave-ftn}.

We are ready to take the continuum limit of the three sums in \eqref{three-sum}. Let us work them out one by one. 
First, we have 
\begin{align}
\begin{split}
A_0^\mathrm{(a)} &\approx -2\pi h^2 |\vec{k}|^2 \int d\left( \frac{1}{\sin\xi} \right) (\sin\xi) J_0( h\alpha \cot\xi) 
        \\
    &= -2\pi h^2 |\vec{k}|^2 \int dx \, x (\sin^2 \xi) J_0(h\alpha x) \,, 
\end{split}
\end{align}
where $x = \cot\xi$, and $\xi$ is now understood as a function of $x$.
Similarly, we find 
\begin{align}
\begin{split}
      A_0^\mathrm{(b)} &\approx -2\pi h^2 |\vec{k}|^2 \int d\left( \frac{1}{\sin\xi} \right) \frac{\cos^2\xi}{\sin\xi} \cos\left(\frac{\pi}{\cos\xi} \right) J_0 ( h\alpha \cot\xi)
      \\
      &= -2\pi h^2 |\vec{k}|^2 \int dx \, x \left[ \cos^2\xi \cos\left(\frac{\pi}{\cos\xi} \right) \right] J_0 ( h\alpha x) \,.
\end{split}
\end{align}
The anti-symmetric part is slightly more interesting. Note that 
\begin{align}
     d^{n+1}_{h,h}(\alpha) - d^{n}_{h,h}(\alpha) &\approx 
     \frac{dx}{dn} \frac{d}{dx} J_0(h\alpha x)
     = - \left( \frac{\alpha}{\cos\xi} \right) J_1(h\alpha x) \,.
\end{align}
Then, the sum becomes 
\begin{align}
\begin{split}
        A_2 &\approx -2\pi i h^2  |\vec{k}|^2  \int d\left( \frac{1}{\sin\xi} \right) \frac{\cos^2\xi}{\sin\xi} \sin\left(\frac{\pi}{\cos\xi} \right) 
    \left( \frac{\alpha}{\cos\xi} \right) J_1(h\alpha \cot\xi) 
    \\
    &= -2\pi ih^2 |\hbar \vec{q}| |\vec{k}| \int dx \, x \left[ \cos\xi \sin\left(\frac{\pi}{\cos\xi} \right) \right] J_1(h\alpha x) \,.
\end{split}
\label{sum-AS}
\end{align}
The argument of the Bessel functions remains finite in the classical limit: 
\begin{align}
    \alpha = 2 \arctan\left(\frac{\hbar|\vec{q}|}{2|\vec{k}|}\right) \approx \frac{\hbar|\vec{q}|}{|\vec{k}|}\,,
    \quad 
    h_\mathrm{cl} = h \,\hbar \,
    \quad 
    \Longrightarrow
    \quad 
    h\alpha x = \frac{h_\mathrm{cl}|\vec{q}|}{|\vec{k}|} \,.
\end{align}

Comparing these with the integrals in \eqref{G-classical-bessel}, 
we find a perfect agreement: 
\begin{align}
      \frac{1}{k} (\hat{k}\cdot\vec{K})|_\mathrm{cl} = \frac{k^2}{\hbar^2} F_0  \,,
        \quad 
     \frac{1}{k}    (\hat{q}\cdot\vec{K})|_\mathrm{cl} = \frac{k^2}{\hbar^2} F_2 \,. 
\end{align} 

\paragraph{Parity odd term} 

Let us proceed to the parity odd term $A_1$ in \eqref{monopole-SS}. 
To begin with, using the spinor notation, we can compute $(\vec{k}_1 \times \vec{k}_3) \cdot\vec{k}_2$:
\begin{align} \label{spinor-cross}
\begin{split}
    (\hat{k}_1 \times \hat{k}_3) \cdot\hat{k}_2 & = \frac{1}{4 i} \mathrm{tr} \left( [\hat{k}_1 \cdot \vec{\sigma}, \hat{k}_3 \cdot \vec{\sigma} ]  (\hat{k}_2 \cdot \vec{\sigma}) \right) 
    \\
    & = 2i [ (k_{3-}|k_{2}^-)(k_{3+}|k_{2}^-)(k_{2-}|k_{1}^-)(k_{2-}|k_{1}^+) 
    \\
    & \qquad \quad 
    - (k_{3+}|k_{2}^+)(k_{3-}|k_{2}^+)(k_{2+}|k_{1}^+)(k_{2+}|k_{1}^-) ]
    \,.
\end{split}
\end{align}
It is convenient to separate $A_1$ into $A_1^-$ and $A_1^+$ which correspond to the $k_{2-}$ term and the $k_{2+}$ term in \eqref{spinor-cross}, respectively. 
The Clebsch-Gordan relation for $A_1^-$ is given in \eqref{wigner-CG1}. The covariant relation is
\begin{align} \label{spinor-CG-}
\begin{split}
    &\quad 2(2n+1) (k'_-|k^-)(k'_-|k^+) \l(k' |k \r)^{n}_{-h,h} 
    \\
    &= k^h_{n+1} \l(k' |k \r)^{n+1}_{-h-1,h} + (k^h_n + l^h_n) \l(k' |k \r)^n_{-h-1,h} 
     + l^h_{n-1} \l(k' |k \r)^{n-1}_{-h-1,h}\,,
\end{split}
\end{align}
where we defined 
\begin{align} 
\begin{split}
    k^h_n &= -\left(1 + \frac{h}{n} \right) \sqrt{(n-h)(n+h+1)}, 
    \\
    l^h_n &= \left( 1-\frac{h}{n+1} \right) \sqrt{(n-h)(n+h+1)} \,.
\end{split}
\end{align}
For $A_1^+$, the covariant version of the Clebsch-Gordan relation \eqref{wigner-CG2} is
\begin{align} \label{spinor-CG+}
\begin{split}
    &\qquad 2(2n+1) (k'_+|k^+)(k'_+|k^-) \l(k' |k \r)^{n}_{-h,h}
    \\
    &= k^{-h}_{n+1} \l(k' |k \r)^{n+1}_{-h+1,h} + (k^{-h}_n + l^{-h}_n) \l(k' |k \r)^n_{-h+1,h} 
    + l^{-h}_{n-1} \l(k' |k \r)^{n-1}_{-h+1,h} \,.
\end{split}
\end{align}
The coefficients on the RHS of \eqref{spinor-CG-} and \eqref{spinor-CG+} are related by $h \rightarrow -h$. 

Using the orthogonality relation, we obtain
\begin{align}
\begin{split}
    A_1^- &= \pi i |\vec{k}_3|^3  \sum_{n=|h|}^{\infty} \frac{1}{2n+3} \l( k_3 | k_1 \r)^{n+1}_{h,h} \times B_n^h  
    \\
    B_n^h &= (k^h_{n+1})^2 + (k^h_{n+1} l^h_{n+1}) + (l^h_{n+1})^2  + (k^h_{n+1} l^h_{n+1}) \cos\pi (l''-l) 
    \\
    &\quad - k^h_{n+1} (k^h_{n+1} + l^h_{n+1}) \cos\pi (l'-l)  - l^h_{n+1} (k^h_{n+1} + l^h_{n+1}) \cos\pi (l''-l') \,,
\end{split}
\end{align}
where we define $l''$ by
\begin{align}
    (l'' + 1/2)^2 = (n + 5/2)^2 - h^2 \,.
\end{align}
A simple sign flip gives $A_1^+$, 
\begin{align}
    A_1^+ = \pi i |\vec{k}_3|^3 & \sum_{n=|h|}^{\infty} \frac{1}{2n+3} \l( k_3 | k_1 \r)^{n+1}_{h,h}  \times B_n^{-h} \,,
\end{align}

Since $A_1 = A_1^- - A_1^+$, only the $h$-odd terms in $B_n^{\pm h}$ survive. Interestingly, $(k^{\pm h}_{n+1} l^{\pm h}_{n+1})$ is completely $h$-even. 
The $h$-odd parts of $(k^{\pm h}_{n+1})^2$ and $(l^{\pm h}_{n+1})^2$ are
\begin{align}
\begin{split}
       (k^{\pm h}_{n+1})^2|_{\text{odd}} &= \pm (2n+3)h\left(1-\frac{h^2}{(n+1)^2} \right), 
    \\
    (l^{\pm h}_{n+1})^2|_{\text{odd}} &= \mp (2n+3)h\left(1-\frac{h^2}{(n+2)^2} \right) \,,
\end{split}
\end{align}
which give
\begin{align}
\begin{split}
    A_1 & = 2 \pi i |\vec{k}_3|^3 h \left[ \sum_{n=|h|}^{\infty}  \left(1-\frac{h^2}{(n+1)^2} \right) \left(1-\cos(\pi \Delta l)\right) \l( k_3 | k_1 \r)^{n+1}_{h,h} \right.
    \\
    &  \left. \hspace{2.7cm} - \sum_{n=|h|+1}^{\infty} \left(1-\frac{h^2}{(n+1)^2} \right)\left(1-\cos(\pi \Delta l)\right) \l(k_3 | k_1 \r)^{n}_{h,h} \right] 
    \\
    & \approx - 2 \pi i |\vec{k}|^3 h^2 \alpha \int dx \, x \, \sin \xi \cos\xi \left[1 -  \cos\left( \frac{\pi}{\cos\xi} \right) \right] \, J_1 \left(h \alpha x \right) \,.
\end{split}
\end{align}
The final result is  
\begin{align}
        \frac{1}{k} (\hat{k}\times \hat{q})\cdot \vec{K}|_\mathrm{cl} = -2 \pi i h^2 \int dx \, x \, c_1 \, J_1 \left(\frac{h_{\mathrm{cl}} |\vec{q}|}{|\vec{k}|} x \right) = \frac{k^2}{\hbar^2} F_1\,,
\end{align}
in perfect agreement with \eqref{c1_c2} and \eqref{G-classical-bessel}.

\subsection{Saddle point approach}

In this subsection, we generalize the Clebsch-Gordan approach 
explained in section~\ref{sec:KMOC-saddle} to the charge-monopole scattering. 
As in the previous subsection, we work out the parity even terms first and proceed to the parity odd term.

\paragraph{Parity even terms}
Let us work out the $\theta$-dependence first. The S-matrix looks like 
\begin{align}
    S(k,\theta)  = 2\pi e^{i\pi h} \sum_{n=|h|}^\infty n_+ e^{2i\delta_n} d^n_{-h,h} (\theta) \,.
\end{align}
The semi-classical formula for the $d$-function, given by CDMST, is 
\begin{align}
\begin{split}
      d^n_{-h,h} (\theta) &\approx (-1)^{n-h} (n_+ c_\theta)^{-1/2} (1-h^2/n_+^2 - c_\theta^2)^{-1/4} 
      \\
      &\qquad \times \cos \left[ n_+ \beta_{n,h}(\theta) -h \gamma_{n,h}(\theta) - \pi/4 \right] \,, 
\end{split}
\end{align}
where the angles $\beta$, $\gamma$ are given by 
\begin{align}
      \sin \left( \frac{1}{2} \beta_{n,h}(\theta) \right) = \frac{ \cos(\theta/2)  }{\cos\xi_{n,h}}  \,,
      &\quad 
      \sin \left( \frac{1}{2} \gamma_{n,h}(\theta) \right)= \frac{\cot(\theta/2)}{\cot\xi_{n,h}} \,, 
\\
     h = n_+ \sin\xi_{n,h} \,,
    &\quad 
    \sqrt{n_+^2 - h^2} = n_+ \cos\xi_{n,h} \,.
\end{align}
The exponential factor from $e^{2i\delta_n} d^n_{-h,h}(\theta)$ 
relevant for the saddle point condition is 
\begin{align}
    \exp \left[ i\left( -\pi \sqrt{n_+^2 -h^2} + n_+\beta_{n,h}(\theta) - h \gamma_{n,h}(\theta) \right)\right] \,.
\end{align}
Differentiating the exponent with respect to $n_+$ while keeping $h$ and $\theta$ fixed, we find the saddle point condition, 
\begin{align}
   - \frac{d}{dn} (2\delta_n) = \frac{\pi}{\cos\xi_{n,h}} = \beta_{n,h}(\theta)
    = 2 \arcsin\left( \frac{\cos(\theta/2)}{\cos\xi_{n,h}} \right) \,,
    \label{saddle-theta-xi}
\end{align}
So, the angle $\beta$ at the saddle point is identified with $\Delta \psi$ in \eqref{Delta-psi}. It also agrees with the classical formula \eqref{scattering-angle-BBCEL-CDMST}.

Evaluating the saddle point integral following \cite{Schwinger:1976fr}, we find 
\begin{align}
\begin{split}
      S(k,\theta) &\approx (-1)^{2h} \frac{2\pi h_\mathrm{cl}}{\hbar}  \left( \frac{1}{\sin^4 \xi} \frac{2 \sin 2\xi}{\sin\theta} \left|\frac{d\xi}{d\theta} \right|\right)^{1/2} e^{iY/\hbar } \,,
      \\
      Y &= -\pi \sqrt{J_*^2 - h_\mathrm{cl}^2} + J_* \beta_* - h_\mathrm{cl} \gamma_* + Y_0  \,.
\end{split}
\label{saddle-CDMST-final}
\end{align}
It is understood that $J_*$, $\beta_*$, $\gamma_*$ are functions of $\theta$ through the saddle point condition \eqref{saddle-theta-xi}. 
The term $Y_0$ is independent of $\theta$ and does not affect the physical observables. 

Now, to use $S^{-1} (i\partial_\theta S)$ in a way similar to \eqref{n-expectation}, 
we need 
\begin{align}
\begin{split}
   \frac{d}{d\theta} \left[ n_+ \beta_{n,h}(\theta) - h \gamma_{n,h}(\theta) \right] 
   &= -\frac{n_+^2 \sin (\theta/2)}{ \sqrt{n_+^2 \sin ^2(\theta/2)-h^2}}
   +\frac{h^2 \csc(\theta/2)}{ \sqrt{n_+^2\sin ^2(\theta/2)-h^2}}
   \\
   &=-\frac{\sqrt{n_+^2 \sin^2 (\theta/2)-h^2}}{\sin (\theta/2)} 
   = -h  \frac{[\sin^2(\theta/2) - \sin^2\xi]^{1/2}}{\sin(\theta/2)\sin\xi} \,.
\end{split}
\label{alpha2-derivation}
\end{align}
It leads to the conclusion that, as far as the $\theta$-dependence is concerned, 
\begin{align}
    S(\vec{k}_2, \vec{k}_1) \approx S(\vec{k}_2, \vec{k}) \exp\left[ \frac{-i\alpha_2(\theta)}{2\sin\theta } \hat{k}_2\cdot \vec{q} \right] \,,
    \quad 
    \alpha_2 = \frac{h_\mathrm{cl} }{k} \frac{[\sin^2(\theta/2) - \sin^2\xi]^{1/2}}{\sin(\theta/2)\sin\xi} \,. 
\end{align}

\paragraph{Parity odd term} 

We expand the three vectors of the KMOC formula 
in the $(\hat{k}, \hat{b}, \hat{k}\times\hat{b})$ basis:
\begin{align}
\begin{split}
     \hat{k}_2 &= \hat{k} \cos\theta  + \hat{b} \sin\theta \cos\phi  + (\hat{k}\times \hat{b}) \sin\theta \sin\phi \,, 
     \\
     \hat{k}_3 &= \hat{k} \cos\theta_q  + \hat{b} \sin\theta_q \cos\phi_q  + (\hat{k}\times \hat{b}) \sin\theta_q \sin\phi_q \,,
     \\
     \hat{k}_1 &= \hat{k} \cos\theta_q  - \hat{b} \sin\theta_q \cos\phi_q  - (\hat{k}\times \hat{b}) \sin\theta_q \sin\phi_q  \,. 
\end{split} 
\end{align}
As always, the angle $\theta_q$ is assumed to be infinitesimal in the classical limit, 
\begin{align}
    \theta_q \approx \frac{\hbar q}{2k} \ll 1 \,.
    \label{theta-infinitesimal}
\end{align}
In contrast, there is no restriction on $\phi_q$. Let $\Delta \phi = \phi_q - \phi$. Geometrically, 
\begin{align}
    \vec{q}\cdot (\hat{k}\times \hat{k}_2) = q \sin\theta \sin(\Delta \phi) \,.
\end{align}

The key step in establishing the ``$\phi$-dependence" of the amplitude is to show that
\begin{align}
\begin{split}
     \frac{\l(k_3|k_2\r)^n_{h,-h} \l(k_2|k_1 \r)^n_{-h,h}}{\l(k|k_2\r)^n_{h,-h} \l(k_2|k\r)^n_{-h,h}} 
    &\approx \exp\left[ i \frac{ h_\mathrm{cl} q}{k} \sin(\Delta\phi) \cot(\theta/2) \right] 
\\
      &= \exp\left[ i \frac{h_\mathrm{cl}}{k} \frac{\cot(\theta/2)}{\sin\theta}  \vec{q}\cdot (\hat{k}\times \hat{k}_2) \right]\,,
\end{split}
\label{phi-dep}
\end{align}
in the classical limit, for an arbitrary $n$. 
We present a short proof of \eqref{phi-dep}. 
Consider the expansion \eqref{base-amp-expansion} of the base amplitude 
specialized to $-b=h=a$: 
\begin{align}
\begin{split}
 \l(k_2 | k_1 \r)^n_{-h,h} 
    &=  \sum_{t= 0}^{ n-|h|}
    \binom{n+h}{t} \binom{n-h}{t}
    \\
    &\qquad \quad \times 
    (k_{2+}|k_{1}^+)^t (k_{2+}|k_{1}^-)^{n-h-t} (k_{2-}|k_{1}^+)^{n+h-t}(k_{2-}|k_{1}^-)^t 
    \,.
\end{split}
\end{align}
In view of the complex conjugation relations \eqref{spinor-cc}, 
\begin{align}
  (k_{2+}|k_{1}^+) = (k_{2-}|k_{1}^-)^*\,, \quad 
   (k_{2+}|k_{1}^-) = - (k_{2-}|k_{1}^+)^* 
   \,,
   \nonumber 
\end{align}
we deduce that the phase of $\l(k_2 | k_1 \r)^n_{-h,h}$ is equal to that of $(k_{2-}|k_{1}^+)^h (k_{2+}|k_{1}^-)^{-h}$. 
The same argument for $\l(k_3| k_2 \r)^n_{h,-h}$ calls for  $(k_{3+}|k_{2}^-)^h (k_{3-}|k_{2}^+)^{-h}$. 
The explicit expressions for the relevant spinor products are  
\begin{align}
\begin{split}
    -(k_{2-}|k_{1}^+) = (k_{2+}|k_{1}^-)^* = e^{-i \Delta \phi/2} c_{\theta_q/2} s_{\theta/2} + e^{i \Delta \phi/2} s_{\theta_q/2} c_{\theta/2} \,,
    \\
    - (k_{3+}|k_{2}^-) = (k_{3-}|k_{2}^+)^* =  e^{i \Delta \phi/2} c_{\theta_q/2} s_{\theta/2} - e^{-i \Delta \phi/2} s_{\theta_q/2} c_{\theta/2} \,.
\end{split}
\end{align}
For the infinitesimal $\theta_q$ in \eqref{theta-infinitesimal}, 
$c_{\theta_q/2}\approx 1$, $s_{\theta_q/2} \approx \theta_q/2$, so that
\begin{align}
\frac{(k_{2-}|k_{1}^+)}{(k_{2+}|k_{1}^-)} & \approx -\frac{1+\frac{\hbar q}{4k} e^{i \Delta \phi} \cot  (\theta/2)}{1+\frac{\hbar q}{4k} e^{-i \Delta \phi} \cot  (\theta/2)} \approx -\left[1+i\frac{\hbar q}{2k} \sin (\Delta \phi) \cot  (\theta/2) \right] 
    \,.
\end{align}
Multiplying it by a similar term from $(k_{3+}|k_{2}^-)(k_{3-}|k_{2}^+)^{-1}$, we find  
\begin{align}
\frac{(k_{3+}|k_{2}^-)(k_{2-}|k_{1}^+)}{(k_{3-}|k_{2}^+)(k_{2+}|k_{1}^-)}  \approx1+i\frac{\hbar q}{k} \sin (\Delta \phi) \cot  (\theta/2) 
    \,.
        \label{phi-dep-3}
\end{align}
In the classical limit where $h\rightarrow \infty$ with $h_\mathrm{cl} = h \hbar$ fixed, we have
\begin{align}
   \lim_{h\rightarrow \infty} (1+\hbar x)^h = \lim_{h\rightarrow \infty} \left(1+\frac{h_\mathrm{cl}}{h} x \right)^h = e^{h_\mathrm{cl}x} \,.
   \label{phi-dep-4}
\end{align}
Combining \eqref{phi-dep-3} and \eqref{phi-dep-4}, we arrive at \eqref{phi-dep}.

\paragraph{KMOC from saddle point} 

We have shown that 
\begin{align}
     &S^\dagger( \vec{k}_3,\vec{k}_2) S(\vec{k}_2, \vec{k}_1) \approx 
    |S(\vec{k}_2,\vec{k})|^2 \exp \left[ -i \frac{\alpha_2}{\sin\theta} (\vec{q} \cdot \hat{k}_2) + i \frac{\alpha_1}{\sin\theta} \vec{q}\cdot (\hat{k}\times \hat{k}_2)  \right] 
\,,
\label{reverse-eng}
\end{align}
where
\begin{align}
    \alpha_1 = \frac{h_\mathrm{cl}}{k} \cot(\theta/2) \,,
    \quad 
    \alpha_2 = \frac{h_\mathrm{cl} }{k} \frac{[\sin^2(\theta/2) - \sin^2\xi]^{1/2}}{\sin(\theta/2)\sin\xi} \,. 
\label{reverse-to-forward}
\end{align}
Here, $\xi$ in $\alpha_2$ is regarded as a function of $\theta$ through the saddle point condition \eqref{saddle-theta-xi}. 

The $q$-integral produces a delta-function similar to \eqref{q-to-delta}:
\begin{align}
    \delta^2\left( \frac{\alpha_2}{\sin\theta} \hat{k}_{2\perp}  - \frac{\alpha_1}{\sin\theta} (\hat{k}\times \hat{k}_{2\perp}) - \vec{b} \right) \,.
\end{align}
The delta function enforces the following relations:
\begin{align}
    \alpha_2 \cos\phi + \alpha_1 \sin\phi = b\,,
    \quad 
     \alpha_2 \sin\phi - \alpha_1 \cos\phi = 0\,. 
     \label{fix-theta-phi}
\end{align}
The relation for the magnitude, $\alpha_1^2 + \alpha_2^2 = b^2$, 
is consistent with the saddle point condition \eqref{saddle-theta-xi} 
as well as the classical relation $h_\mathrm{cl}/k b = \tan\xi$. 
Assuming the relation between $\theta$ and $\xi$, we can determine $\phi$ by 
\begin{align}
    \tan\phi = \frac{\alpha_1}{\alpha_2} = \sin\xi \tan\left(\frac{\pi}{2\cos\xi} \right) = \frac{c_1}{c_2}  \,, 
\label{phi-reverse-copy}
\end{align}
in perfect agreement with \eqref{c1_c2}. 

It remains to show that the Jacobian produced by the $\hat{k}'$-integral, 
analogous to \eqref{k-prime-scalar}, 
cancels against $|S|^2$ in \eqref{reverse-eng}. 
The Jacobian is given by 
\begin{align}
    J(\theta) = \frac{1}{(2\pi)^2} \frac{\sin\theta}{|\alpha_1(\theta) \alpha_1'(\theta) + \alpha_2(\theta) \alpha_2'(\theta)|} \,.
    \label{jacobian-monopole}
\end{align}
A straightforward computation shows that 
it indeed cancels against $|S|^2$ from \eqref{saddle-CDMST-final}.

\section{Discussion} 

We have shown how the KMOC formulation and the partial wave expansion together reproduce the classical impulse in the purely conservative sector. We presented the partial wave expansion using spinor products as base amplitudes to facilitate the incorporation of arbitrary spin of the particles as well as the extra ``field" angular momentum  of the charge-monopole system. 

There are a few directions in which this work can be extended. The most obvious one is to apply our methods to gravitational binary systems. 
Although we only took examples from electromagnetic interaction in this paper, the generalization to gravitational interaction should be straightforward. 

In particular, it would be interesting to see how the spin-dependent interactions are encoded in the partial wave amplitudes. At the leading order in Newton's constant $G$, the spin-dependent Hamiltonian has been computed to all orders in spin \cite{Guevara:2018wpp,Chung:2018kqs,Guevara:2019fsj,Chung:2019duq,Chung:2020rrz}, 
which leads to a simple formula for the classical deflection of the spin directions. 
Reproducing the same result from the partial wave expansion will serve as a consistency check, and clarify how the massive spinor helicity variable \cite{Arkani-Hamed:2017jhn} is related to the spinor basis used in this paper. 
See, e.g., \cite{Cristofoli:2021jas} for related discussions. 

The most important assumption throughout this paper is that the partial wave phase converges to the radial action in the classical limit. In the probe limit, this assumption can be verified rather easily to all orders in coupling (proportional to the inverse of the total angular momentum). 
Away from the probe limit, one has to rely on perturbation theory in coupling, and the emergence of the radial action is not too obvious. The exponential representation of \cite{Damgaard:2021ipf} seems to be a particularly promising way to retrieve the radial action systematically. It would be interesting to further develop the exponential representation to incorporate the spin-dependence.

One of the main strengths of the KMOC formulation is that the conservative sector and the radiative sector can be treated in a uniform way. 
The analysis of this paper was strictly restricted to the conservative sector. Generalizing it to include the radiation is another important open problem. 
See, e.g., \cite{Cristofoli:2021vyo,Saketh:2021sri} for related discussions. 

It would be also interesting to study the charge-monopole scattering for an arbitrary mass ratio. The naive perturbation theory in appendix~\ref{app:pert-charge-momopole} of this paper, while conceptually straightforward, lacks computational efficiency already at the next-to-leading order. More efficient methods such as \cite{Mogull:2020sak} could be adopted to the charge-monopole scattering. 
The Coulomb scattering away from the probe limit was discussed in detail in \cite{Bern:2021xze}.

In the probe limit, the charge-monopole scattering problem can be solved to all orders as shown in \cite{Schwinger:1976fr,Boulware:1976tv,Kol:2021jjc} and reproduced in this paper. Going beyond the probe limit, even perturbatively, in a manifestly Lorentz covariant way appears to be an open problem. 
As pointed out by Weinberg \cite{Weinberg:1965rz} and Zwanziger \cite{Zwanziger:1970hk}, a Feynman-like propagator for the charge-monopole system carries a gauge-dependent ``Dirac-string" 4-vector. 
It would be interesting to rewrite the amplitude, at least the part surviving the classical limit, in a gauge invariant way. 
The pairwise helicity variables introduced by \cite{Csaki:2020inw} looks promising in this regard. The authors of \cite{Csaki:2020inw} showed that 
the charge-monopole two-particle phase space cannot be a simple product of the two single-particle phase space. 
This fact could serve as a guiding principle when we try to find 
a manifestly Lorentz covariant expression for 
the S-matrix $S(\vec{k}',\vec{k})$ we computed in this paper. 
See \cite{Terning:2018lsv,Terning:2018udc,Caron-Huot:2018ape, Moynihan:2020gxj} for some recent discussions on charge-monopole scattering. 

\section*{Acknowledgements}

This work was supported in part by the National Research Foundation of Korea grant NRF-2019R1A2C2084608. 
We are grateful to Joon-Hwi Kim, Jung-Wook Kim, Kanghoon Lee and Sungjay Lee 
for discussions and comments on the manuscript. 
S.M. would like to thank SINP, Kolkata; TIFR, Mumbai and the organizers of `QCD Meets Gravity' conference for hospitality while this work was in progress.

\newpage
\appendix 

\section{Wigner and Bessel functions}
\label{app:wave-ftn}

\paragraph{Wigner d-matrices} 

We spell out our convention for the Wigner $d$-matrices, $d^n_{a,b}(\theta)$. We will not use the full $D(\alpha,\beta,\gamma)$ matrices. Instead, we will discuss the phase factors separately if necessary. 
We begin with the original definition of Wigner, 
\begin{align}
    d^n_{a,b}(\theta) = \langle n, a | e^{-i\theta J_y} | n,b \rangle \,.
\end{align}
For the simplest non-trivial example, we set $n=1/2$ and $J_y = \sigma_y/2$: 
\begin{align}
    d^{1/2}(\theta) = 
        \begin{pmatrix}
        d^{1/2}_{1/2,1/2} & d^{1/2}_{1/2,-1/2} \\
        d^{1/2}_{-1/2,1/2} & d^{1/2}_{-1/2,-1/2}
    \end{pmatrix} 
    =
    \begin{pmatrix}
        c_{\theta/2} & -s_{\theta/2} \\
        s_{\theta/2} & c_{\theta/2}
    \end{pmatrix} \,.
\end{align}
The matrix elements for higher $n$ can be understood from the ``total symmetrization" construction. 
For general $d^n_{b,a}$, Wigner gave a constructive formula, 
\begin{align}
\begin{split}
      d_{b, a}^n(\theta) &= \left[\left(n+b\right) !\left(n-b\right) !(n+a) !(n-a) !\right]^{\frac{1}{2}} 
      \\
      &\quad \times \sum_{s=s_0}^{s_1} \frac{(-1)^{b-a+s}\left(c_{\theta/2}\right)^{2 n+a-b-2 s}\left(s_{\theta/2}\right)^{b-a+2 s}}{(n+a-s) ! s !\left(b-a+s\right) !\left(n-b-s\right) !} \,.  
\end{split}
\label{d-expansion-Wigner} 
\end{align}
The range of $s$ is such that the factorials in the sum are non-negative: 
\begin{align}
   s_0 =\max \left(0, m-m^{\prime}\right), \quad s_1 =\min \left(j+m, j-m^{\prime}\right) \,.
\end{align}
The corresponding formula for our spinor product reads
\begin{align}
\begin{split}
  \l(k_2 | k_1 \r)^n_{b,a}
    &=  \sum_{t}
    N_{n,b,a,t}
    (k_{2+}|k_{1}^+)^t (k_{2+}|k_{1}^-)^{n+b-t} (k_{2-}|k_{1}^+)^{n+a-t}(k_{2-}|k_{1}^-)^{t-a-b} \,,
    \\
    &\quad N_{n,b,a,t} = 
    \sqrt{ \binom{n+b}{t} \binom{n-b}{t-a-b}
    \binom{n+a}{t} \binom{n-a}{t-a-b} }
    \,,
\end{split}
\label{base-amp-expansion}
\end{align}
where $t$ is restricted such that all exponents in the sum are non-negative.
It reduces to Wigner's formula \eqref{d-expansion-Wigner} if we adjust the angles such that $(k_{2\pm}| k_{1\pm}) = d^{1/2}_{\pm 1/2,\pm 1/2}$, 
and rename the summation variable as $t= n+a-s$. 

It is sometimes useful to express $d$-functions in terms 
of Jacobi polynomials \cite{QTAM:1988}, 
\begin{align}
    d_{b,a}^n(\theta)=\xi_{b,a}\left[\frac{s !(s+\mu+\nu) !}{(s+\mu) !(s+\nu) !}\right]^{\frac{1}{2}}\left(s_{\theta/2}\right)^\mu\left(c_{\theta/2}\right)^\nu P_s^{(\mu, \nu)}(c_\theta) \,,
\end{align}
where $\xi_{b,a} = 1$ (if $a\ge b$) or $ (-1)^{a-b}$ (if $a<b$), and 
\begin{align}
    \mu = \left|b-a\right|, \quad 
    \nu = \left|b+a\right|, \quad 
    s = n-\frac{1}{2}(\mu+\nu) .
\end{align}

\paragraph{Clebsch-Gordan}

To derive the KMOC formula for the charge-monopole scattering, 
we need some Clebsch-Gordan relations. 
For the parity even terms, all we need is 
\begin{align}
c_\theta \, d^n_{-h,h} (\theta) =  
\frac{(n+1)^2-h^2 }{(n+1) (2 n+1)} d^{n+1}_{-h,h} (\theta)
-\frac{h^2 }{n (n+1)} d^n_{-h,h} (\theta)
+\frac{n^2-h^2 }{n (2 n+1)} d^{n-1}_{-h,h} (\theta) \,.
\label{wigner-CG}
\end{align}
For the parity odd terms, we need two closely related ones. One is 
\begin{align}
\begin{split}
     s_\theta \, d^n_{-h,h}(\theta) &= - \frac{(n+1-h) \sqrt{(n+1+h) (n+2-h)}}{(n+1) (2 n+1)} d^{n+1}_{-h+1,h} (\theta)
     \\
     &\quad 
+\frac{h \sqrt{(n+h) (n+1-h)}}{n (n+1)} d^n_{-h+1,h} (\theta)
\\
&\quad 
+\frac{(n+h)\sqrt{(n-1+h) (n-h)}}{n (2 n+1)} d^{n-1}_{-h+1,h} (\theta) \,.
\end{split}
\label{wigner-CG1}
\end{align}
The other is 
\begin{align}
\begin{split}
     s_\theta \, d^n_{-h,h}(\theta) &=  
     +\frac{(n+1+h) \sqrt{(n+1-h) (n+2+h)}}{(n+1) (2 n+1)} d^{n+1}_{-h,h+1} (\theta)
     \\
     &\quad 
+\frac{h \sqrt{(n-h) (n+1+h)}}{n (n+1)} d^n_{-h,h+1} (\theta)
\\
&\quad 
-\frac{(n-h)\sqrt{(n-1-h) (n+h)}}{n (2 n+1)} d^{n-1}_{-h,h+1} (\theta) \,.
\end{split}
\label{wigner-CG2}
\end{align}
The two relations, \eqref{wigner-CG1} and \eqref{wigner-CG2}, can be mapped to each other by 
\begin{align}
    d_{a,b}^n = (-1)^{b-a} d_{b,a}^n = d_{-b,-a}^n\,.
\end{align}

\paragraph{Bessel functions}
 
We quote a few well known properties of spherical Bessel functions. 
The orthogonality relation reads
\begin{align}
    \int_0^{\infty}  j_l(k' r) j_l(k r) r^2 d r = \frac{1}{4 k^2} \hat{\delta}(k'-k)\,.
    \label{ortho-sb}
\end{align}
The asymptotic form for $x \gg 1$ is 
\begin{align}
    j_l(x) \approx \frac{1}{x} \sin\left( x- \frac{\pi l}{2} \right) + \frac{l(l+1)}{2x^2} \cos\left( x- \frac{\pi l}{2} \right) \,.
\end{align}
In the absence of the monopole, an exact decomposition 
is known for the plane-wave:
\begin{align}
    e^{ikz} = e^{ikr \cos\theta} = \sum_{l=0}^\infty (2l+1) e^{i \pi l/2} j_l(kr) P_l(\cos\theta)\,.
\end{align}

\paragraph{Wigner vs Bessel for small angles} 

In the main text, when we take the classical limit, we approximate Wigner d-functions by Bessel functions. Here, we review a derivation of the approximation. 
Our starting point is the eigenvalue equation, $\vec{\mathcal{J}}^2 D^n_{h_1,h_2} = n(n+1) D^n_{h_1,h_2}$, which leads to a 
differential equation for $w^n_{h_1,h_2}(\theta) \equiv (\sin\theta)^{1/2} d^n_{h_1,h_2}(\theta)$:
\begin{align}
\left[ \frac{d^2}{d \theta^2}+ n_+^2-\frac{(h_1-h_2)^2-1/4}{4 \sin ^2 (\theta/2)}-\frac{(h_1+h_2)^2-1/4}{4 \cos ^2(\theta/2)}\right] w^n_{h_1,h_2}(\theta) = 0 \,.   
\end{align}
Consider the limit where $h = (h_1+ h_2)/2 \gg 1$ with $a = h_2-h_1$ kept small ($|a|\sim 1$). At the same time, we take the small angle limit  ($\theta \ll 1$). 
In this double limit, we have 
\begin{align}
    n_+^2  -\frac{(h_1+h_2)^2-1/4}{4 \cos ^2(\theta/2)} \approx n_+^2 - h^2 \,,
    \quad 
    \frac{(h_1-h_2)^2-1/4}{4 \sin ^2 (\theta/2)} 
    \approx \frac{a^2-1/4}{\theta^2} \,,
\end{align}
so that the differential equation is approximated by 
\begin{align}
\left[ \frac{d^2}{d \theta^2} + l_+^2 -\frac{a^2-1/4}{\theta^2} \right] w^n_{h_1,h_2}(\theta) = 0 \,, 
\quad 
l_+ = \sqrt{n_+^2 - h^2} \,.
\end{align} 
If we scale $w^n_{h_1,h_2}$ by $\sqrt{\theta}$ as 
\begin{align}
    y^n_{h,a}(\theta) = \theta^{-1/2} w^n_{h_1,h_2} (\theta)
    = \left(\frac{\sin\theta}{\theta}\right)^{1/2} d^n_{h_1,h_2}(\theta) \,,
\end{align}
the differential equation for $y^n_{h,a}$ in the double limit is given by 
\begin{align}
    \left[ \theta^2 \frac{d^2}{d\theta^2} + \theta \frac{d}{d\theta} +(l_+^2\theta^2-a^2) \right] y^n_{h,a}(\theta) = 0 \,.
\end{align}
It is precisely the defining equation for the Bessel function $J_a(l_+ \theta)$. 

It remains to check the normalization of $y^n_{h,a}(\theta)$.
The answer is simply
\begin{align}
    d^n_{h_1,h_2}(\theta) \approx \left(\frac{\theta}{\sin\theta}\right)^{1/2} J_a(l_+\theta) \approx J_a(l_+\theta) \,.
\end{align}
For $a= 0$, the normalization is easily fixed by the fact that 
$d^n_{h,h}(0)=1 = J_0(0)$ for any $n\ge |h|$. 
For a non-zero $a$, one way to check the normalization is to use the action of the ladder operators on the d-matrices. 
For example, it is known that 
\begin{align}
\begin{split}
    \left(- \frac{d}{d\theta} - \frac{h_1}{\sin\theta}  +  \frac{h_2 \cos\theta}{\sin\theta} \right) d^n_{h_1,h_2}(\theta) 
   = \sqrt{n(n+1) - h_2(h_2+1)} d^n_{h_1,h_2+1}(\theta) \,.
\end{split}
\end{align}
In the double limit, this relation is approximated by 
\begin{align}
   \frac{1}{l_+} \left( - \frac{d}{d\theta} + \frac{a}{\theta} \right) d^n_{h_1,h_2}(\theta)
   \approx  d^n_{h_1,h_2+1}(\theta)\,.
\end{align} 
It matches with the well known relation for Bessel function, 
\begin{align}
     \left( - \frac{d}{dx} + \frac{a}{x} \right) J_a(x) 
   = J_{a+1}(x)\,, 
   \quad 
   d^n_{h_1,h_2}(\theta) \approx J_{h_2-h_1}(l_+\theta) \,.
\end{align}

\section{Classical perturbation for charge-monopole scattering} 
\label{app:pert-charge-momopole}

We follow section~6.1 of \cite{Kosower:2018adc} to carry out a classical perturbation theory for the charge-monopole scattering in a manifestly Lorentz covariant way. It can be done beyond the probe limit, 
but we will focus on the probe limit where we can make comparison 
to the results in the main text. 

\paragraph{Leading order} 
The 0-th order trajectories are taken to be straight lines:
\begin{align}
    x_1(\tau_1) = b + u_1 \tau_1 \,,
    \quad 
    x_2(\tau_2) = u_2 \tau_2 \,. 
\end{align}
Here, $x_{1,2}$, $u_{1,2}$ and $b$ are 4-vectors while $\tau_{1,2}$ are scalars. 
For the Coulomb scattering, the field strength of particle 2 is given by 
\begin{align}
    F_2^{\mu\nu}(x) = -i e_2 \int \hat{d}^4 q\, \hat{\delta}(q\cdot u_2) e^{iq\cdot x} \frac{q^\mu u_2^\nu - u_2^\mu q^\nu}{q^2}  
    \,.
    \qquad \mbox{(Coulomb)}
\end{align}
Applying an electromagnetic duality transformation, 
\begin{align}
    F^{\mu\nu} \;\rightarrow \; \frac{1}{2} \epsilon^{\mu\nu}{}_{\rho\sigma} F^{\rho\sigma} \,,
    \quad 
    e \;\rightarrow \; g \,, 
\end{align}
we obtain the field strength when particle 2 is a monopole, 
\begin{align}
    F_2^{\mu\nu}(x) = -i g_2 \int \hat{d}^4 k\, \hat{\delta}(k\cdot u_2) e^{ik\cdot x} \frac{\epsilon^{\mu\nu}{}_{\rho\sigma} k^\rho u_2^\sigma }{k^2}  \,.
    \qquad \mbox{(Monopole)}
    \label{FF-monopole}
\end{align}
In what follows, to avoid clutter, we introduce shorthand notations, 
\begin{align}
\hat{d}^3 k_\perp  &\equiv \hat{d}^4 k\,  \hat{\delta}\left(k \cdot u_2\right) \,,
\quad 
\hat{d}^2 k_\perp  \equiv \hat{d}^4 k\, \hat{\delta}\left(k \cdot u_1\right) \hat{\delta}\left(k \cdot u_2\right) \,.
\end{align}

Inserting \eqref{FF-monopole} into the Lorentz force law for particle 1, 
\begin{align}
    \frac{dp_1^\mu}{d\tau} = -e_1 F_2^{\mu\nu}(x_1) (u_1)_\nu \,, 
\end{align}
we obtain 
\begin{align} \label{classic_force}
   \frac{d p_1^\mu}{d \tau} = -  i e_1 g_2 \int \hat{d}^4 k \, \hat{\delta}\left(k \cdot u_2\right) e^{i k \cdot\left(b+u_1 \tau_1\right)} \frac{\epsilon^\mu{}_{\nu\rho\sigma} u_1^\nu u_2^\rho k^\sigma }{k^2}  \,.
\end{align}
The leading order impulse turns out to be 
\begin{align}
\Delta p_1^{\mu,(1)} &= \int_{-\infty}^{\infty} d \tau \frac{d p_1^\mu}{d \tau} 
= - i e_1 g_2 \int \hat{d}^2 k_\perp 
e^{i k \cdot b} \frac{\epsilon^\mu{}_{\nu\rho\sigma} u_1^\nu u_2^\rho k^\sigma}{k^2} 
\,.
\label{monopole-impulse}
\end{align}

\paragraph{Evaluation} 

Let us evaluate \eqref{monopole-impulse}. 
To maintain Lorentz covariance as much as we can, 
we use $u_1$, $u_2$, $b$ and $\varepsilon^\mu(u_1,u_2,b) = \varepsilon^\mu{}_{\nu\rho\sigma} u_1^\nu u_2^\rho b^\sigma$ as a 
non-orthonormal basis:
\begin{align}
\begin{split}
     V^\mu &= \frac{1}{(u_1\cdot u_2)^2-1} \left[ (V \cdot u_1 + (u_1\cdot u_2)V\cdot u_2) u_1^\mu + (V \cdot u_2+ (u_1\cdot u_2)V\cdot u_1) u_2^\mu \right]
     \\
     &\quad - (V \cdot \hat{b}) \hat{b}^\mu  -  (V\cdot \hat{\varepsilon})  \hat{\varepsilon}^\mu \,, 
     \\
     \hat{b}^\mu &= b^\mu/|b| \,,
     \quad 
     \varepsilon^\mu(u_1,u_2,b) = \varepsilon^\mu{}_{\nu\rho\sigma} u_1^\nu u_2^\rho b^\sigma\,,
     \quad 
     \hat{\varepsilon}^\mu = \frac{\varepsilon^\mu(u_1,u_2,b)}{|\varepsilon^\mu(u_1,u_2,b)|} \,.
\end{split}
\label{cov-basis}
\end{align}
A short computation shows that 
\begin{align}
    \Delta p_1^{\mu,(1)} \cdot u_1 = \Delta p_1^{\mu,(1)} \cdot u_2 = \Delta p_1^{\mu,(1)} \cdot b = 0 \,, 
\end{align} 
and that 
\begin{align}
    \Delta p_1^{\mu,(1)} = \frac{e_1 g_2}{2\pi |b| } \hat{\varepsilon}^\mu \,.
    \label{impulse-kmo-monopole}
\end{align} 
In the probe limit, \eqref{impulse-kmo-monopole} 
agrees with \eqref{impulse-bbcel-leading} as expected. 

\paragraph{Sub-leading order}
We continue to the 2nd order in the probe limit. 
\begin{align}
    \Delta p_1^{\mu, (2)} = \int_{-\infty}^{\infty} d\tau \, \Delta^{(2)}\left(\frac{dp_1^\mu}{d\tau}\right)
\end{align}
Recall that the force is given by $\eqref{classic_force}$, 
\begin{align}
\begin{split}
    \Delta^{(2)}\left(\frac{dp_1^\mu}{d\tau}\right) = -  i e_1 g_2 \int \hat{d}^3 l_\perp e^{i l \cdot\left(b+u_1 \tau_1\right)}\epsilon^{\mu }{}_{\nu\rho\sigma} \left(\frac{d}{d \tau} \Delta^{(1)} x_1^\nu + i(l \cdot \Delta^{(1)} x_1) u_1^\nu \right) \frac{ u_2^\rho l^\sigma }{l^2} \,.
\end{split}    
\end{align}
Here, $\Delta^{(1)} x_1$ is the leading order correction on particle 1's trajectory. 
Using the leading order results in the previous section,
\begin{align}
    \Delta^{(1)} x^\mu_1 = i\frac{e_1 g_2}{m} \int \hat{d}^3 k_\perp e^{i k \cdot (b+u_1 \tau)} \frac{\epsilon^{\mu}{}_{\nu \rho \sigma} u_1^\nu u_2^\rho k^\sigma}{k^2 (k \cdot u_1 -i \epsilon)^2} \,.
\end{align}
Relabelling $q=k+l$, we obtain the desired formula for the sub-leading impulse:
\begin{align}
\begin{split}
    \Delta p_1^{\mu, (2)} &=  i \frac{(e_1g_2)^2}{m_1} \int \hat{d}^2 q_\perp e^{i q \cdot b} \, R(u_1,u_2,q) \,,
    \\ 
     R(u_1,u_2,q) &=  
     \int \hat{d}^3 l_\perp
     \left[ -\frac{[u_1^\mu+u_2^\mu (u_1 \cdot u_2)] (q -l) \cdot l - (q-l)^\mu (l \cdot u_1 )}{l^2 (q-l)^2 (l \cdot u_1 + i \epsilon)} \right. \\
    & \hspace{6cm} \left. + \frac{\epsilon^{\mu}{}_{\nu \rho \sigma} u_1^\nu u_2^\rho l^\sigma \, \epsilon(l, u_1, u_2, q)}{l^2 (q-l)^2 (l \cdot u_1 + i \epsilon)^2} \right] \,.
\end{split}
\label{classical-impulse-NLO}
\end{align}
We can split this vector integral to a few scalar integrals using the basis \eqref{cov-basis}. 
Projections onto the $u_2$, $u_1$ directions give $\Delta p \cdot u_2 = 0$ and 
\begin{align} 
    \Delta p \cdot u_1 &= i \frac{(e_1g_2)^2}{m_1} \int \hat{d}^2 q_\perp e^{i q \cdot b} 
    \int \hat{d}^3 l_\perp
     \frac{\left[1-(u_1 \cdot u_2)^2 \right] (q -l) \cdot l - (l \cdot u_1)^2}{l^2 (q-l)^2 (l \cdot u_1 + i \epsilon)} \,.
\end{align}
The component in the $b$ direction is 
\begin{align} 
\begin{split}
    &\Delta p \cdot b = i \frac{(e_1g_2)^2}{m_1} \int \hat{d}^2 q_\perp e^{i q \cdot b}
    \\ 
    &\qquad \quad \times  \int \hat{d}^3 l_\perp
     \left[ \frac{b\cdot (q-l) (l \cdot u_1)}{l^2 (q-l)^2 (l \cdot u_1 + i \epsilon)}  + \frac{\epsilon(b, u_1, u_2, l) \, \epsilon(l, u_1, u_2, q)}{l^2 (q-l)^2 (l \cdot u_1 + i \epsilon)^2} \right]\,, 
    \\
    &\epsilon(\Delta p, u_1,u_2,b) = i \frac{(e_1g_2)^2}{m_1} \int \hat{d}^2 q_\perp e^{i q \cdot b}
    \\ 
    &\qquad \quad \times  \int  \hat{d}^3 l_\perp
     \left[ \frac{ - \epsilon(q-l,u_1,u_2,b) (l \cdot u_1)}{l^2 (q-l)^2 (l \cdot u_1 + i \epsilon)} \right. \\
    & \hspace{6cm} \left. + \frac{\epsilon_{\mu}(u_1, u_2, b)  \epsilon^{\mu}(u_1, u_2, l) \, \epsilon(l, u_1, u_2, q)}{l^2 (q-l)^2 (l \cdot u_1 + i \epsilon)^2} \right] \,.
\end{split}
\end{align}
The integrals can be done by elementary methods. 
The simpler ones give 
\begin{align}
\begin{split}
    &\Delta p \cdot u_1 = i \frac{(e_1g_2)^2}{m_1} \int \hat{d}^2 q_\perp e^{i q \cdot b}
    \int \hat{d}^3 l_\perp
     \frac{\left[1-(u_1 \cdot u_2)^2 \right] (q -l) \cdot l - (l \cdot u_1)^2}{l^2 (q-l)^2 (l \cdot u_1 + i \epsilon)}
     \\
     & \qquad \quad =  -2 m \kappa^2 \left[ (u_1 \cdot u_2)^2-1 \right] \,,
    \\
    &\epsilon(\Delta p, u_1,u_2,b) = 0
    \,.
\end{split}
\end{align}
In the probe limit, $(\Delta p\cdot u_1)$ agrees perfectly with \eqref{impulse-bbcel-sub-leading}.

The most challenging one is $(\Delta p\cdot b)$, for which we evaluate two integrals separately:
\begin{align}
\begin{split}
     (\Delta p \cdot b)_{\text{A}} 
     &= i \frac{(e_1g_2)^2}{m_1} \int \hat{d}^2 q_\perp e^{i q \cdot b}
     \int \hat{d}^4 l \, \hat{\delta}(l \cdot u_2)
     \frac{ b\cdot (q-l)}{l^2 (q-l)^2}
    \\ 
    &= \frac{\pi}{2}  \kappa^2 mv |b|  \,,
     \\
    (\Delta p \cdot b)_{\text{B}} &= i \frac{(e_1g_2)^2}{m_1} \int \hat{d}^2 q_\perp e^{i q \cdot b}
     \int \hat{d}^3 l_\perp
      \frac{\epsilon(b, u_1, u_2, l) \, \epsilon(l, u_1, u_2, q)}{l^2 (q-l)^2 (l \cdot u_1 + i \epsilon)^2}
    \\
     &= -\pi \kappa^2 mv |b|
    \,.
\end{split}
\end{align}
The two terms add up to give 
\begin{align}
  (\Delta p \cdot b) =   (\Delta p \cdot b)_{\text{A}} + (\Delta p \cdot b)_{\text{B}} = -\frac{\pi}{2} \kappa^2 mv |b|\,.
\end{align}
In the probe limit, $(\Delta p\cdot b)$ agrees perfectly with \eqref{impulse-bbcel-sub-leading}.


\bibliographystyle{unsrt}
\bibliography{reference}
 
\end{document}